\documentclass[useAMS,usenatbib,usegraphicx]{mn2e}

\def\lea{\mathrel{<\kern-1.0em\lower0.9ex\hbox{$\sim$}}}
\def\gea{\mathrel{>\kern-1.0em\lower0.9ex\hbox{$\sim$}}}

\title[M32 RR Lyrae Variables]{An Ancient Metal-Poor Population in M32, and Halo Satellite 
Accretion in M31, Identified by RR Lyrae Stars\footnote{Based on observations taken with the NASA/ESA 
Hubble Space Telescope, obtained at the Space Telescope Science Telescope.}}
\author[A. Sarajedini et al.]{Ata Sarajedini$^{1}$\thanks{E-mail:
ata@astro.ufl.edu}, S. -C. Yang$^{1}$, A. Monachesi$^{2}$, Tod R. Lauer$^{3}$, and S. C. Trager$^{4}$\\
$^{1}$University of Florida, Department of Astronomy, 211 Bryant Space Science Center, Gainesville, FL, 32611 USA\\
$^{2}$Department of Astronomy, University of Michigan, 830 Dennison Building, 500 Church Street, Ann Arbor, MI 48109, USA\\
$^{3}$National Optical Astronomy Observatory, P.O. Box 26732, Tucson, AZ 85726, USA\\
$^{4}$Kapteyn Astronomical Institute, P.O. Box 800, 9700 AV Groningen, The Netherlands}

\begin{document}

\date{}

\pagerange{\pageref{firstpage}--\pageref{lastpage}} \pubyear{2002}

\maketitle

\label{firstpage}

\begin{abstract}
 We present time-series photometry of two fields near
M32 using archival observations from the Advanced Camera for Surveys
Wide Field Channel onboard the Hubble Space Telescope. One field is
centered about 2 arcmin from M32 while the other is located 15 arcmin
to the southeast of M31. The imaging covers a time baseline sufficient
for the identification and characterization of a total number of 1139
RR Lyrae variables of which 821 are ab-type and 318 are c-type. In the
field near M32, we find a radial gradient in the density of RR Lyraes
relative to the center of M32. This gradient is consistent with the
surface brightness profile of M32 suggesting that a significant number
of the RR Lyraes in this region belong to M32. This provides further
confirmation that M32 contains an ancient stellar population formed
around the same time as the oldest population in M31 and the Milky
Way. The RR Lyrae stars in M32 exhibit a mean metal abundance of
$\langle[\mathrm{Fe/H}]\rangle\approx-1.42\pm 0.02$, which is
$\approx$ 15 times lower than the metal abundance of the overall M32
stellar population. Moreover, the abundance of RR Lyrae stars
normalized to the luminosity of M32 in the field analyzed further
indicates that the ancient metal-poor population in M32 represents
only a very minor component of this galaxy, consistent with the 1\% to
4.5\% in mass inferred from the CMD analysis of Monachesi et al. We
also find that the measured reddening of the RR Lyrae stars is
consistent with M32 containing little or no dust. In the other field,
we find unprecedented evidence for two populations of RR Lyraes in M31
as shown by two distinct sequences among the ab-type variables in the
Bailey Diagram. When interpreted in terms of metal abundance, one
population exhibits a peak at $[\mathrm{Fe/H}]\approx-1.3$ and the
other is at $[\mathrm{Fe/H}]\approx-1.9$. One possible interpretation
of this result is that the more metal-rich population represents the
dominant M31 halo, while the metal-poorer group could be a disrupted
dwarf satellite galaxy orbiting M31.  If true, this represents a
further indication that the formation of the M31 spheroid has been
significantly influenced by the merger and accretion of dwarf galaxy
satellites. 
\end{abstract}


\section[]{Introduction}

\subsection{RR Lyraes and their use in unveiling stellar populations}

The class of pulsating stars known as RR Lyrae variables are located
at the intersection of the instability strip and the horizontal branch in the
Hertzsprung Russell Diagram (Smith 1995). There are three principal types 
of RR Lyrae variables; those pulsating in the
fundamental mode exhibit sawtooth-like light curves and are referred to as
ab-type or RR0 variables. The first overtone pulsators generally show sine-curve
shaped light curves, have shorter periods and typically lower amplitudes than the
ab-types, and are referred to as c-type or RR1 variables. Lastly, RR Lyraes that pulsate
in both the fundamental and first overtone modes (i.e. double mode pulsators) 
carry the d-type moniker.

The astrophysical utility of RR Lyraes to investigate a number of key
questions in the areas of stellar populations and galaxy formation is
well documented. For example, because of their low masses
($\approx0.7\,M_\odot$, Smith 1995), the mere presence of RR Lyrae
stars in a stellar population suggests an old age ($\gea10$ Gyr) for
the system. As such, one does not need to obtain deep photometry
beyond the old main sequence turnoff in order to establish the
presence of an old population.

The periods and amplitudes of the ab-type RR Lyrae stars ($P_{ab}$)
are related to their metallicities. Using data on Milky Way field RR
Lyraes from Layden (2005, private communication) and those in the
Large Magellanic Cloud, Sarajedini et al.\ (2006) and Alcock et
al.\ (2000), respectively, present relations that correlate the
periods and amplitudes of RR Lyraes with their metal abundances.  Once
the metallicities of the RR Lyraes are determined, their absolute
magnitudes can be calculated using relations between [Fe/H] and the
absolute magnitude of the RR Lyraes [$M_V(\mathrm{RR})$], which is
then used to estimate the distance to their parent population, be it a
star cluster or a galaxy.  Lastly, ab-type RR Lyraes are also useful
for calculating the line-of-sight reddening.  The minimum light colors
of these stars are largely independent of their other properties as
shown by Guldenschuh et al.\ (2005) and Kunder et al.\ (2010) making
the determination of reddening a simple one-step process.

Thus far, we have presented examples of how RR Lyrae variables can be
powerful probes of the systems in which they reside - star clusters or
among the field populations of galaxies. It is for this reason that
studying them in Local Group galaxies like M31 and M32 provides
valuable insights into the properties of these systems.

\subsection{RR Lyrae variables in M31}

The study of RR Lyraes in M31 has a rich history dating back to the
seminal work of Pritchet \& van den Bergh (1987) and culminating in
the most recent papers by Sarajedini et al.\ (2009) and Jeffrey et
al.\ (2011). Sarajedini et al.\ (2009, hereafter S09) present Hubble
Space Telescope (HST) Advanced Camera for Surveys (ACS) observations
for two fields in the range of 4 to 6 kpc from the center of M31. A
total of 681 RR Lyraes (555 ab-type and 126 c-type) were identified in
the two fields. A mean metal abundance of $\mathrm{[Fe/H]}\approx-1.5$
was determined using the periods and amplitudes of these stars.

Jeffery et al.\ (2011) is a continuation of the work of Brown et
al.\ (2004) and presents high quality light curves for RR Lyraes in 5
fields around M31. These include a halo field at 21 kpc, and two halo
fields at 35 kpc. In addition, there is a field coincident with one of
the streams in the vicinity of M31 and one that covers a disk region
about 26 kpc from the center along the major axis of the galaxy.
Besides identifying a number of RR Lyrae variables in these fields,
Jeffery et al.\ (2011) also compare three methods for the
determination of RR Lyrae metallicities.  We note in passing also that
Bernard et al.\ (2012) present RR Lyraes in two M31 fields along the
major axis based on HST/ACS observations. These are located in the
disk warp and in the outer disk.

\subsection{RR Lyrae variables in M32}

M32 is an intriguing galaxy, considered to be the prototype of the
compact elliptical galaxies. In view of its importance in
understanding these objects, its stellar populations have been
extensively studied (e.g., Rose 1985, Freedman 1992, Grillmair et
al. 1996, Worthey 2004, Trager et al. 2000, Rose et
al. 2005). Recently, using the High Resolution Channel (HRC) on ACS,
Monachesi et al.\ (2011, 2012) have analyzed the stellar populations of
M32 in a field $\sim 2\arcmin$ away from its nucleus. They have
constructed the deepest optical color magnitude diagram (CMD) of M32
ever and derived its star formation history (SFH). They found that
M32 has had an extended SFH and is composed of two main dominant
populations: a 2--5 Gyr old, metal rich population and a population
older than 5 Gyr, with slightly subsolar metallicity.  Their study
moreover indicates that a significant contribution from stars older
than 10 Gyr is not expected, although there are claims from
spectroscopic analysis that such a population should exist in M32
(e.g., Coelho et al. 2009). Given the extreme crowding in their field,
Monachesi et al.\ (2011) were unable to detect the oldest main sequence
turnoff (MSTO) of M32 and thus a census of the oldest stars in M32 is
still missing. RR Lyrae stars are therefore a direct indicator
of the presence of an ancient population in M32. Alonso-Garcia et
al.\ (2004) used the Wide Field Planetary Camera 2 (WFPC2) onboard HST
to image a field $\approx 3.5$ arcmin to the east of M32.  Comparing
the variable star content of this field with that of a control field
that samples the M31 field away from M32, they claim to have
identified RR Lyraes that belong to M32, although this result remains
quite uncertain, since their data suffer from strong photometric and
temporal incompleteness. A marginal detection of RR Lyrae stars in M32
was presented by Fiorentino et al.\ (2010). They used the same HRC
field of HST/ACS as Monachesi et al.\ (2011) and identified 17 RR Lyrae
variables. Using a Bayesian analysis, they suggest that M32 contains
RR Lyrae stars and therefore possesses an old ($\gea 10$ Gyr)
population. However, the small field of view of ACS/HRC and the strong
contamination from M31 makes this detection also very uncertain.

As we were preparing this paper, we became aware of the most recent
work from Fiorentino et al.\ (2012), which presents their reduction and
analysis of a subset of the data presented herein. They studied the RR
Lyraes in the M32 field and investigated their spatial distribution as
well as their periods and metallicities.  We will highlight their work
in more detail within the relevant sections of the present paper, but
for the moment, we note that their results are largely in agreement
with ours.

\subsection{This paper}

In the present work, we utilize archival wide field channel (WFC)
HST/ACS images of two fields in the vicinity of M32 in order to search
for RR Lyraes. The ACS/WFC fields have a wider spatial coverage of
M32's stellar populations over a range of higher surface brightnesses
than the previous WFPC2 and ACS/HRC fields studied. Our primary task
is to identify a substantial population of RR Lyraes belonging to M32
that would confirm the presence of a truly ancient, metal-poor
population in this galaxy, and characterize their properties so that
we can use them to study their parent galaxy.  Furthermore, we probe
the RR Lyraes in the spheroid of M31 in order to better understand
their properties: their metallicities and Oosterhoff types.

The paper is organized as follows. Sections 2 and 3 describe our
observations and data reduction including the artificial star
experiments we use to characterize our photometric
incompleteness. Section 4 provides a description of the techniques
used to identify and characterize the RR Lyraes in our sample. We also
detail the simulations used to assess any biases present in our
variability data. The results and discussion as they pertain to M32 
are in Section 5, where we
investigate the membership of the RR Lyraes in our field closest to
M32. Section 6 presents the results and discussion of the RR Lyraes 
that belong to M31, where we show unprecedented evidence for
two sequences of ab-type RR Lyraes in the Bailey Diagram.
Finally, the conclusions of the present work are given in Section 7.

\section[]{Observations}

The observations used in the present study were obtained with 
HST/ACS/WFC as part of program GO-9392 (PI: Mateo).  The details of
this program have already been fully described by Rudenko et
al.\ (2009) and the observing log is shown in Table 1. In summary, two
fields were observed; the first one, which we designate `M32' in
Fig. 1, was centered southeast of M32 covering a radial distance range
of 1 to 4.5 arcmin from the center of M32. Figure 2 shows a reproduction of
the M32 field observed in the F606W filter. The second field was meant
to be a control field (designated 'Control' in Fig. 1) located roughly
5 arcmin northeast of M32 in order to monitor the M31 background field
in the vicinity of M32. The M32 field was observed for a total of
$\sim5$ hours in the F606W filter and $\sim8$ hours in F814W
alternating between the two filters and yielding 18 exposures in each
filter. These were taken on 2004 November 24--25 and 2004 December
10. Similarly, the Control field was observed on 2004 December 20--22
for $\sim5$ and $\sim7.5$ hours in F606W and F814W, respectively,
again alternating between filters and providing 18 exposures per
filter. Figure 3 shows raw unphased light curves for two of our RR
Lyrae variables in order to illustrate the cadence of the
observations.

It is important to note that, as we establish
below, the properties of the M31 RR Lyraes in the Control field are significantly
different from those in the M32 field. As such, we did not make use of the Control
field in the manner it was originally intended - to gauge the level of contamination 
of M32 stars by the M31 background. Instead, we have determined the background
density using selected data from the M32 field as well as the observations
presented by S09. Section 5.2 provides more details on the
procedure we followed.

\begin{figure}
\includegraphics[width=85mm]{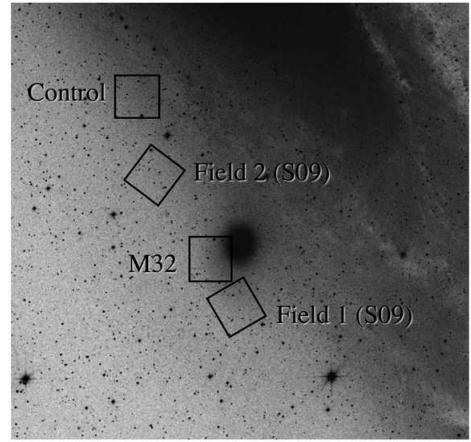}
\caption{The location of our ACS fields (M32 and Control) along with
  the two fields from S09 overplotted on a digitized sky survey image
  in the region of M31. The dwarf elliptical galaxy M32 is near the
  center of the image. North is up and east is to the left.}
\end{figure}

\begin{figure}
\includegraphics[width=80mm]{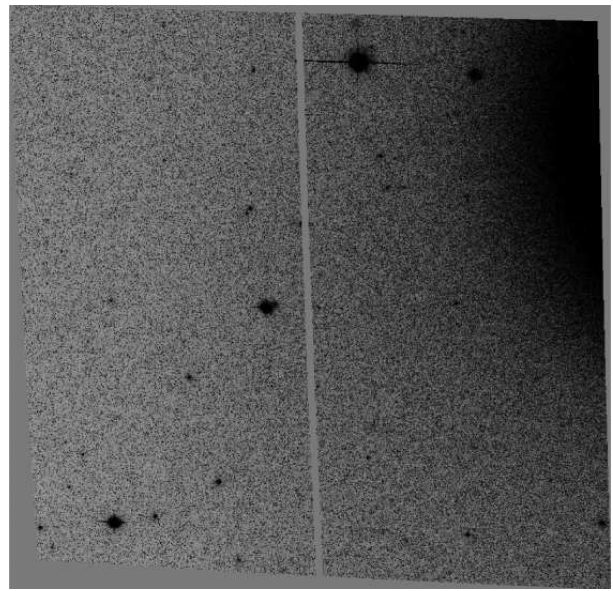}
\caption{The F606W filter drizzled ACS image of our M32 field.
North is up and east is to the left.}
\end{figure}

\begin{figure}
\includegraphics[width=100mm]{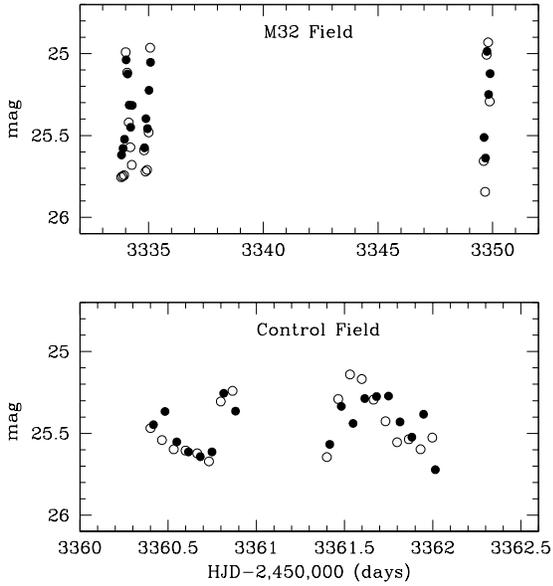}
\caption{Raw unphased light curves for two of the RR Lyraes in our sample
illustrating the cadence of the observations in each field. The filled
circles are the F606W filter data while the open circles are in F814W.}
\end{figure}

\begin{table*}
 \centering
 \begin{minipage}{150mm}
  \caption{Observing Log.}
  \begin{tabular}{@{}ccccccc@{}}
  \hline
   Field &  R. A. (J2000) & Dec (J2000) &  Starting Date & Data Sets & Filter & Exp Time \\
 \hline
M32 & 00 42 55.92  & +40 50 50.4 &  2004 Nov 24 &  J8F101, J8F103, J8F105 & F606W &  2 x 900s, 1 x 960s, 15 x 1000s  \\
   &    &    &      &     & F814W    & 1 x 1520s, 3 x 1570s, 14 x 1580s   \\
   &    &    &      &     &   &    \\
Control & 00 43 28.65   &   +41 03 43.8  &  2004 Dec 20    &   J8F102, J8F104, J8F106  & F606W    &   2 x 900s, 1 x 960s, 15 x 1000s    \\
   &   &    &     &     & F814W    &  1 x 1520s, 3 x 1570s, 14 x 1580s   \\
\hline
\end{tabular}
\end{minipage}
\end{table*}

\section{Data Reduction}

\subsection{Program Frames}

The 72 ACS/WFC images were measured using the DAOPHOT/ALLSTAR/ALLFRAME
(Stetson 1987; 1994) suite of crowded field photometry programs. The
procedure we followed is identical to that used by Sarajedini et
al.\ (2009, hereafter S09).  In summary, once a master coordinate file
was constructed for each image and high precision coordinate
transformations between the images established, ALLFRAME was used to
fit a high signal-to-noise point spread function to all detected
profiles on each of the 36 images in each field. The measurements on
each of the individual frames were then matched and only stars
appearing on all 36 images were kept.

The standardization procedure of the individual magnitudes is
identical to that used by Sarajedini et al.\ (2006). More specifically,
we have made use of the Reiss \& Mack (2004) prescription to account
for the effects of charge transfer efficiency along with the Sirianni
et al.\ (2005) calibration equations and coefficients to transform our
instrumental photometry to the ground-based $VI$ system.  Each of our
magnitude measurements is affected by three sources of systematic
error: the uncertainty in the aperture corrections ($\pm0.02$ mag),
the error in the correction to infinite aperture for the F606W
($\pm0.00$ mag) and F814W ($\pm0.001$ mag) filters, and the error in
the ground system VI zero point ($\pm0.05$ mag, Sirianni et al. 2005).

\subsection{Artificial Star Experiments}

In order to gauge the degree of photometric completeness on our
program frames, we have performed a series of artificial star
experiments. Beginning with a synthetic color-magnitude diagram (CMD)
generated using IAC-STAR (Aparicio \& Gallart 2004) that replicates
the appearance of the actual CMD produced from our observations, we
selected 50,000 artificial stars. These were placed on the WFC2
images, which is the portion of the field closer to M32, 5,000 stars
at a time, in a grid pattern with random intra-pixel positions. The
WFC1 region of the field, which is less crowded than WFC2, received
35,000 artificial stars. The resultant images were photometered using
the same procedure as the original images. The overall recovery rates
for the artificial stars were 87\% for the WFC1 images and 81\% for
WFC2.

The recovery rates were used to devise a relation between completeness
fraction and radial distance from the center of M32. Since we are
interested in correcting the red clump stars and RR Lyraes for
photometric incompleteness (see below), we limited this relation to
stars in the magnitude range of the horizontal branch.

\section{Characterization of the Variable Stars and Simulations}

We use the technique of Yang et al.\ (2010) and Yang \& Sarajedini
(2012) to identify and characterize variable stars in the images.  We
start by searching all of the variable star candidates located in a
range of $V$ magnitude ($24.5 < V < 26$).  Stars within this $V$
magnitude range were evaluated using a reduced $\chi^2_{VI}$ defined
by the following equation :

\begin{displaymath}
\chi^2_{VI} = \frac{1}{N_V + N_I} \times
\Bigg[\sum_{i=1}^{N_V} \frac{(V_i - \overline V)^2}{\sigma_i^2} +
\sum_{i=1}^{N_I} \frac{(I_i - \overline I)^2}{\sigma_i^2}\Bigg]
\end{displaymath}

\noindent Any anomalous data points that deviated from the mean
magnitude by $\pm3\sigma$ were excluded from the $\chi^2_{VI}$
calculation for each star. Then, we selected only those stars with
$\chi^2_{VI}$ values greater than $3.0$ as potential variable
candidates.

The next step involved running our template light curve fitting
routine, RRFIT (Yang \& Sarajedini 2011) on the time series photometry
of these variable star candidates in order to detect and characterize
RR Lyrae stars. RRFIT uses 25 unique light curve templates (23 RRab type
and 2 RRc type variables). It searches over specific period and
amplitude ranges and calculates the $\chi^2$ difference between the
observed data and each light curve template. It then determines the
optimal light curve parameters such as period, amplitude, maximum
epoch, and mean magnitude from the best-fit template (i.e. the
template that minimizes the $\chi^2$ value).

After applying the above technique, we have identified and
characterized 509 RR Lyrae variables (375 ab-type and 134 c-type) in
the M32 field and 630 RR Lyraes (446 ab-type and 184 c-type) in the
Control field.  By way of comparison, Fiorentino et al.\ (2012) found
416 RR Lyrae variables (314 ab-type and 102 c-type) in the same M32
field analyzed herein. Figure 4 shows a representative sample of our
light curves. Tables 2--9 list the basic properties of each of these
RR Lyrae variables.

In order to investigate the presence of any biases in our
period-finding algorithm, we have performed extensive simulations
wherein we produce $\sim1000$ synthetic RR Lyrae light curves with a
range of ab-type and c-type periods and amplitudes. We then input
these raw synthetic light curves into RRFIT in order to determine
their properties and compare them with their known values. The results
are shown in Figs. 5 and 6, where we plot the difference between the
input and output periods ($\Delta$P) as a function of input period.
In order to estimate the errors of the individual RR Lyrae periods, we
performed the following statistical test introduced in Yang et
al.\ (2010) and also used in Yang \& Sarajedini (2012). From the
artificial RR Lyrae lists, we randomly draw the same number of artificial
RR Lyrae stars as the observed RR Lyraes in each field [M32 : 375 (RRab) and 134
  (RRc); Control : 446 (RRab) and 184 (RRc)] to calculate an average
$\Delta$P value for each sample drawn. We consider the spread of a
$\Delta$P distribution as the estimate of the error. To increase its
statistical significance, we iterated this sampling 10,000 times. The
average spread ($\langle\sigma\rangle$) of 10,000 samples provides a
realistic error for our period measurements. We applied the same
method to calculate errors in the $V$-band amplitude. The resulting
values of the errors are shown in Table 10. Also displayed in this
table is the fraction of simulations where the recovered period
differed from the input period by less than $\pm0.05\,\mathrm{d}$ and
$\pm0.1\,\mathrm{d}$.


We note that the simulations reveal little or no aliasing among the RR
Lyraes periods of the Control field. Even in the M32 field, where
Fig. 5 does show the signature of aliasing, the number of stars in
the aliased `bands' is a small fraction of the total number of
simulations run as demonstrated by the values in Table 10. In fact, we
found no significant biases in our determination of the mean periods
and amplitudes as shown by the relatively small errors listed in Table
10. Hence, we assume that period aliasing does not significantly affect
the results in the remainder of our analysis, which is mainly focused
on the mean periods, mean amplitudes, and results derived from them.

\begin{table*}
 \begin{minipage}{300mm}
  \caption{ab-type RR Lyraes in M32 WFC1 Field}
\begin{tabular}{@{}cccccccccc@{}}
  \hline
   Star ID &  R. A. (J2000) & Dec (J2000) &  $\langle$V$\rangle$ & $\langle$I$\rangle$ & (V--I)$_{min}$ & $\langle$V--I$\rangle$ & Period (days) & V Amplitude & I Amplitude \\
 \hline
  69090 & 0 42 54.89 & +40 49 36.18 &  25.666 &  25.001 &   0.831 &   0.684 &   0.5632 &   0.7199 &   0.4600\\
  55439 & 0 42 55.13 & +40 50 50.89 &  25.271 &  24.610 &   0.778 &   0.673 &   0.6231 &   0.5623 &   0.3681\\
  61847 & 0 42 55.29 & +40 51 45.40 &  25.399 &  24.790 &   0.637 &   0.610 &   0.5599 &   0.4055 &   0.3626\\
  50225 & 0 42 55.30 & +40 50 45.96 &  25.217 &  24.628 &   0.788 &   0.622 &   0.5827 &   0.8878 &   0.5310\\
  54773 & 0 42 55.33 & +40 50 39.33 &  25.118 &  24.606 &   0.701 &   0.546 &   0.5175 &   0.9801 &   0.6470\\
  48015 & 0 42 55.35 & +40 50 18.52 &  25.086 &  24.581 &   0.711 &   0.555 &   0.5197 &   1.0407 &   0.6680\\
  67388 & 0 42 55.41 & +40 50 \phantom{0}5.04 &  25.341 &  24.893 &   0.745 &   0.521 &   0.4624 &   1.4403 &   0.9278\\
\hline
\end{tabular}
\end{minipage}
\end{table*}

\begin{table*}
 \begin{minipage}{300mm}
  \caption{c-type RR Lyraes in M32 WFC1 Field}
  \begin{tabular}{@{}cccccccccc@{}}
  \hline
   Star ID &  R. A. (J2000) & Dec (J2000) &  $\langle$V$\rangle$ & $\langle$I$\rangle$ & (V--I)$_{min}$ & $\langle$V--I$\rangle$ & Period (days) & V Amplitude & I Amplitude \\
 \hline
  55904 & 0 42 55.28 & +40 50 41.80 &  25.223 &  24.707 &   0.625 &   0.525 &   0.3358 &   0.3853 &   0.2382\\
  62969 & 0 42 55.37 & +40 52 28.13 &  25.351 &  24.976 &   0.478 &   0.384 &   0.3177 &   0.4386 &   0.2929\\
  57604 & 0 42 55.69 & +40 49 58.02 &  25.223 &  24.741 &   0.620 &   0.492 &   0.3629 &   0.4065 &   0.2225\\
  47179 & 0 42 55.79 & +40 49 32.05 &  25.108 &  24.696 &   0.490 &   0.417 &   0.3836 &   0.3765 &   0.2668\\
  62947 & 0 42 55.94 & +40 51 42.68 &  25.323 &  24.958 &   0.467 &   0.373 &   0.3050 &   0.4364 &   0.2886\\
  62331 & 0 42 56.01 & +40 51 36.24 &  25.292 &  24.984 &   0.407 &   0.317 &   0.2760 &   0.4680 &   0.3216\\
  46671 & 0 42 56.12 & +40 51 \phantom{0}9.32 &  25.090 &  24.667 &   0.519 &   0.431 &   0.3759 &   0.3684 &   0.2302\\
\hline
\end{tabular}
\end{minipage}
\end{table*}

\begin{table*}
 \begin{minipage}{300mm}
  \caption{ab-type RR Lyraes in M32 WFC2 Field}
  \begin{tabular}{@{}cccccccccc@{}}
  \hline
   Star ID &  R. A. (J2000) & Dec (J2000) &  $\langle$V$\rangle$ & $\langle$I$\rangle$ & (V--I)$_{min}$ & $\langle$V--I$\rangle$ & Period (days) & V Amplitude & I Amplitude \\
 \hline
 161985 & 0 42 45.81 & +40 50 29.49 &  25.407 &  24.774 &   0.768 &   0.648 &   0.5709 &   0.6127 &   0.4081\\
 124879 & 0 42 45.86 & +40 49 15.40 &  25.238 &  24.708 &   0.620 &   0.540 &   0.5760 &   0.6285 &   0.4898\\
 119671 & 0 42 46.02 & +40 50 55.64 &  25.299 &  24.872 &   0.507 &   0.437 &   0.5869 &   0.9324 &   0.8322\\
 145978 & 0 42 46.03 & +40 50 40.41 &  25.160 &  24.752 &   0.592 &   0.445 &   0.4481 &   1.0279 &   0.7026\\
 131710 & 0 42 46.04 & +40 50 38.09 &  25.260 &  24.904 &   0.375 &   0.358 &   0.7845 &   0.4529 &   0.4138\\
 160482 & 0 42 46.11 & +40 50 32.97 &  25.471 &  24.800 &   0.834 &   0.696 &   0.5253 &   0.8770 &   0.6202\\
 170827 & 0 42 46.14 & +40 50 23.19 &  25.412 &  24.920 &   0.673 &   0.523 &   0.4562 &   0.9407 &   0.6043\\
\hline
\end{tabular}
\end{minipage}
\end{table*}

\begin{table*}
 \begin{minipage}{300mm}
  \caption{c-type RR Lyraes in M32 WFC2 Field}
  \begin{tabular}{@{}cccccccccc@{}}
  \hline
   Star ID &  R. A. (J2000) & Dec (J2000) &  $\langle$V$\rangle$ & $\langle$I$\rangle$ & (V--I)$_{min}$ & $\langle$V--I$\rangle$ & Period (days) & V Amplitude & I Amplitude \\
 \hline
 163374 & 0 42 45.64 & +40 49 42.33 &  25.324 &  24.961 &   0.403 &   0.366 &   0.3409 &   0.3612 &   0.2910\\
 140009 & 0 42 45.76 & +40 49 35.26 &  25.256 &  24.761 &   0.592 &   0.502 &   0.3209 &   0.3690 &   0.2131\\
  81800 & 0 42 46.41 & +40 52 \phantom{0}8.19 &  25.053 &  24.267 &   0.846 &   0.793 &   0.3529 &   0.5520 &   0.4156\\
 121539 & 0 42 46.52 & +40 49 12.69 &  25.193 &  24.692 &   0.524 &   0.502 &   0.3424 &   0.3508 &   0.3187\\
 123696 & 0 42 46.54 & +40 50 55.11 &  25.218 &  24.673 &   0.587 &   0.548 &   0.2986 &   0.3276 &   0.2707\\
 178394 & 0 42 46.57 & +40 50 29.67 &  25.539 &  25.159 &   0.433 &   0.385 &   0.2841 &   0.4103 &   0.3262\\
 157733 & 0 42 46.63 & +40 50 22.54 &  25.236 &  24.847 &   0.439 &   0.393 &   0.3090 &   0.4810 &   0.3972\\
\hline
\end{tabular}
\end{minipage}
\end{table*}

\begin{table*}
 \begin{minipage}{300mm}
  \caption{ab-type RR Lyraes in Control WFC1 Field}
  \begin{tabular}{@{}cccccccccc@{}}
  \hline
   Star ID &  R. A. (J2000) & Dec (J2000) &  $\langle$V$\rangle$ & $\langle$I$\rangle$ & (V--I)$_{min}$ & $\langle$V--I$\rangle$ & Period (days) & V Amplitude & I Amplitude \\
 \hline
  59041 & 0 43 27.88 & +41  4 26.80 &  25.055 &  24.599 &   0.555 &   0.472 &   0.5134 &   0.7479 &   0.5663\\
  80706 & 0 43 27.89 & +41  2 14.93 &  25.348 &  24.858 &   0.579 &   0.503 &   0.5276 &   0.6531 &   0.5057\\
  75357 & 0 43 27.93 & +41  4 54.96 &  25.370 &  24.817 &   0.632 &   0.559 &   0.6035 &   0.4618 &   0.3593\\
  74424 & 0 43 27.95 & +41  4 26.71 &  25.264 &  24.756 &   0.648 &   0.527 &   0.5516 &   0.8250 &   0.6131\\
  63686 & 0 43 27.95 & +41  4 \phantom{0}0.55 &  25.216 &  24.628 &   0.809 &   0.626 &   0.5834 &   1.0540 &   0.7148\\
  90791 & 0 43 28.00 & +41  3 \phantom{0}5.46 &  25.685 &  24.996 &   0.682 &   0.688 &   0.5469 &   0.3511 &   0.3626\\
  68832 & 0 43 28.16 & +41  2 45.91 &  25.193 &  24.721 &   0.614 &   0.493 &   0.5600 &   0.8514 &   0.6097\\
\hline
\end{tabular}
\end{minipage}
\end{table*}

\begin{table*}
 \begin{minipage}{300mm}
  \caption{c-type RR Lyraes in Control WFC1 Field}
  \begin{tabular}{@{}cccccccccc@{}}
  \hline
   Star ID &  R. A. (J2000) & Dec (J2000) &  $\langle$V$\rangle$ & $\langle$I$\rangle$ & (V--I)$_{min}$ & $\langle$V--I$\rangle$ & Period (days) & V Amplitude & I Amplitude \\
 \hline
  65042 & 0 43 27.91 & +41  4 41.46 &  25.173 &  24.839 &   0.421 &   0.340 &   0.3229 &   0.3431 &   0.2088\\
  47031 & 0 43 27.94 & +41  2 \phantom{0}9.82 &  25.031 &  24.629 &   0.475 &   0.407 &   0.3033 &   0.3518 &   0.2256\\
  57030 & 0 43 28.09 & +41  3 25.20 &  25.223 &  24.700 &   0.576 &   0.528 &   0.3240 &   0.4322 &   0.3611\\
  70608 & 0 43 28.29 & +41  4 \phantom{0}3.26 &  25.249 &  24.853 &   0.496 &   0.404 &   0.3141 &   0.4313 &   0.2836\\
  76098 & 0 43 28.40 & +41  4 \phantom{0}0.59 &  25.440 &  24.845 &   0.663 &   0.600 &   0.2849 &   0.3845 &   0.2912\\
  71804 & 0 43 28.42 & +41  4 20.76 &  25.190 &  24.706 &   0.504 &   0.485 &   0.4015 &   0.2650 &   0.2377\\
  80532 & 0 43 28.52 & +41  4 46.32 &  25.366 &  25.063 &   0.373 &   0.310 &   0.2814 &   0.4860 &   0.3638\\
\hline
\end{tabular}
\end{minipage}
\end{table*}

\begin{table*}
 \begin{minipage}{300mm}
  \caption{ab-type RR Lyraes in Control WFC2 Field}
  \begin{tabular}{@{}cccccccccc@{}}
  \hline
   Star ID &  R. A. (J2000) & Dec (J2000) &  $\langle$V$\rangle$ & $\langle$I$\rangle$ & (V--I)$_{min}$ & $\langle$V--I$\rangle$ & Period (days) & V Amplitude & I Amplitude \\
 \hline
  83936 & 0 43 18.42 & +41  5 \phantom{0}8.04 &  25.314 &  24.737 &   0.773 &   0.602 &   0.5473 &   0.7587 &   0.4600\\
  82752 & 0 43 18.43 & +41  5 14.88 &  25.255 &  24.672 &   0.866 &   0.630 &   0.5550 &   1.0333 &   0.5900\\
  99546 & 0 43 18.43 & +41  4 33.19 &  25.372 &  24.855 &   0.671 &   0.536 &   0.4587 &   0.6070 &   0.3421\\
  89174 & 0 43 18.51 & +41  4 24.55 &  25.305 &  24.752 &   0.581 &   0.557 &   0.5740 &   0.5056 &   0.4549\\
  60763 & 0 43 18.53 & +41  4 47.27 &  25.208 &  24.532 &   0.725 &   0.679 &   0.6127 &   0.4068 &   0.3328\\
 102771 & 0 43 18.64 & +41  2 60.00 &  25.269 &  24.842 &   0.620 &   0.472 &   0.4868 &   1.2036 &   0.8786\\
 118814 & 0 43 18.68 & +41  2 \phantom{0}6.56 &  25.490 &  25.161 &   0.406 &   0.342 &   0.4551 &   0.7728 &   0.6389\\
\hline
\end{tabular}
\end{minipage}
\end{table*}

\begin{table*}
 \begin{minipage}{300mm}
  \caption{c-type RR Lyraes in Control WFC2 Field}
  \begin{tabular}{@{}cccccccccc@{}}
  \hline
   Star ID &  R. A. (J2000) & Dec (J2000) &  $\langle$V$\rangle$ & $\langle$I$\rangle$ & (V--I)$_{min}$ & $\langle$V--I$\rangle$ & Period (days) & V Amplitude & I Amplitude \\
 \hline
  65950 & 0 43 18.49 & +41  4 40.08 &  25.099 &  24.640 &   0.568 &   0.465 &   0.3337 &   0.3478 &   0.2000\\
 103817 & 0 43 18.57 & +41  3 \phantom{0}7.61 &  25.355 &  24.972 &   0.418 &   0.386 &   0.3460 &   0.3824 &   0.3252\\
  91215 & 0 43 18.78 & +41  2 24.17 &  25.231 &  24.824 &   0.432 &   0.409 &   0.3439 &   0.3137 &   0.2731\\
  93640 & 0 43 18.83 & +41  4 51.84 &  25.347 &  24.807 &   0.634 &   0.547 &   0.3330 &   0.3698 &   0.2415\\
  53982 & 0 43 18.83 & +41  4 \phantom{0}6.12 &  24.948 &  24.471 &   0.522 &   0.481 &   0.3692 &   0.2962 &   0.2375\\
  91682 & 0 43 19.17 & +41  4 34.53 &  25.287 &  24.879 &   0.416 &   0.408 &   0.3207 &   0.3438 &   0.3294\\
 100571 & 0 43 19.19 & +41  5 11.30 &  25.350 &  24.906 &   0.455 &   0.445 &   0.3204 &   0.3032 &   0.2883\\
\hline
\end{tabular}
\end{minipage}
\end{table*}

\begin{table*}
\centering
\begin{minipage}{140mm}
 \caption{Results of the Light Curve Simulations}
 \begin{tabular}{@{}cccccc@{}}
 \hline
  Field &  RR Lyrae type &  Period Error & Amplitude Error &Percent Within $\pm$0.05d & Percent Within $\pm$0.1d \\
\hline
M32 & RRab  & $\pm$ 0.0014 &  $\pm$ 0.0084 & 83\% & 96\% \\
M32 & RRc  & $\pm$ 0.0104 &  $\pm$ 0.0061 & 79\% &  81\% \\
Control & RRab  & $\pm$ 0.0008 &  $\pm$ 0.0055  & 94\% & 98\%  \\
Control & RRc  & $\pm$ 0.0044 &  $\pm$ 0.0046 & 91\% &  94\%  \\
\hline
\end{tabular}
\end{minipage}
\end{table*}

\begin{figure}
\includegraphics[width=90mm]{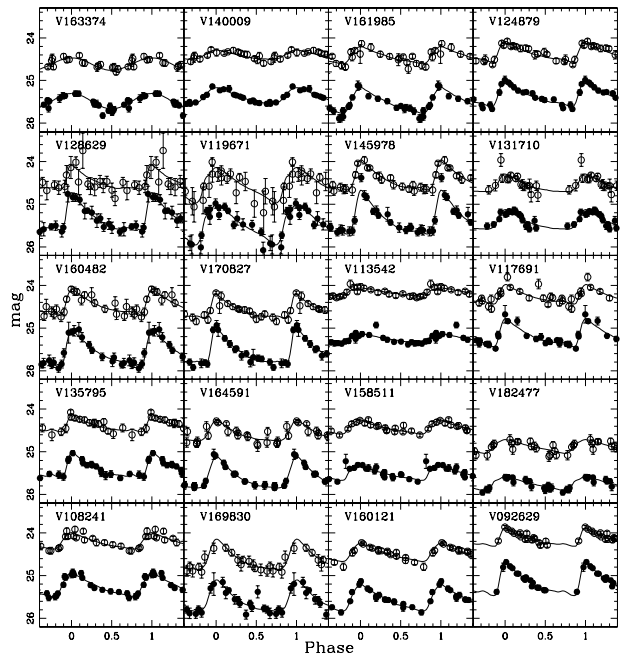}
\caption{Example light curves of some of the RR Lyraes in our sample. The filled
circles are the F606W filter data while the open circles are in F814W. The latter
have been adjusted brighter by 1 mag for illustrative purposes. }
\end{figure}

\begin{figure}
\includegraphics[width=90mm]{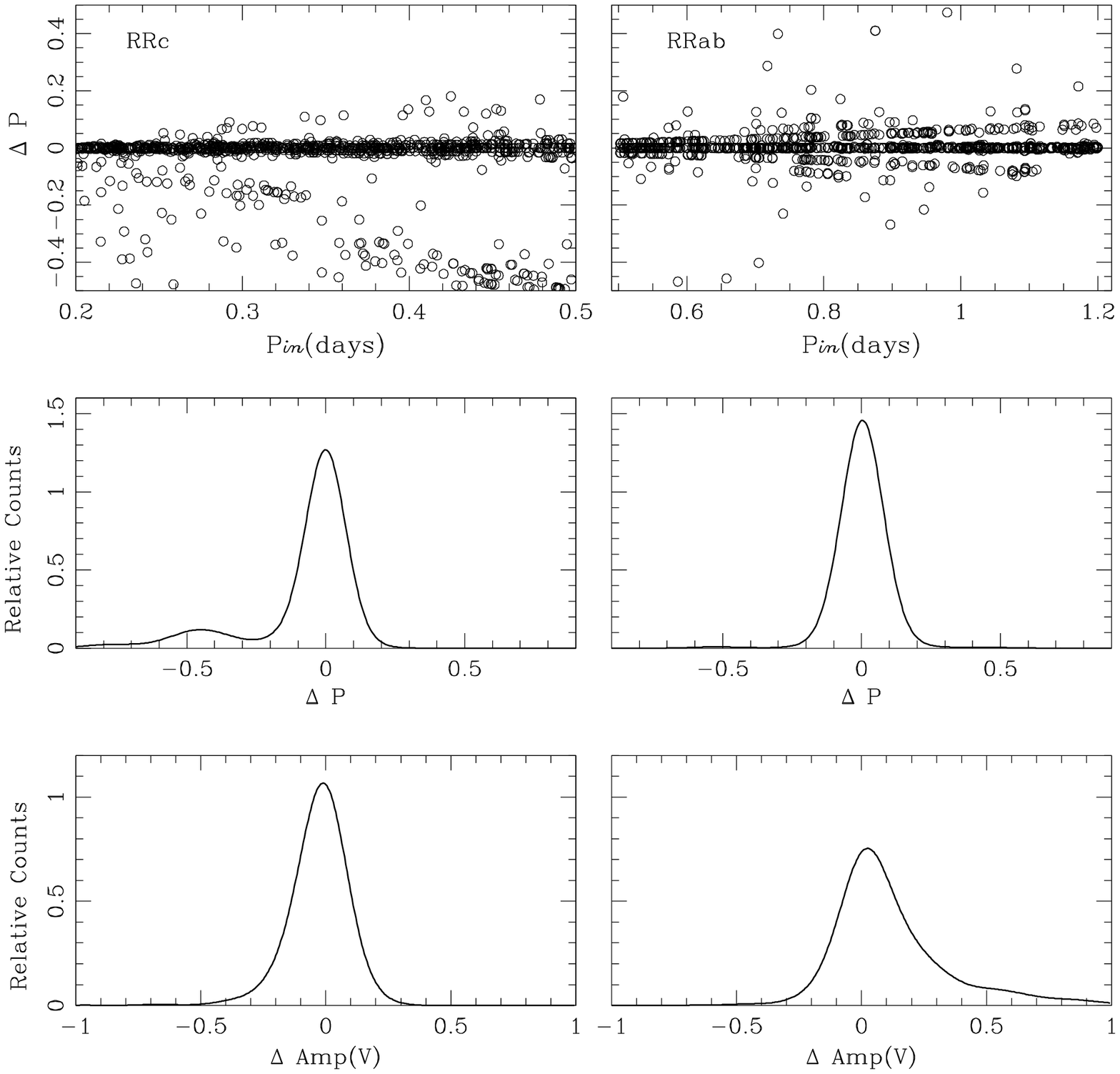}
\caption{The results of our RR Lyrae light curve simulations using an observing
cadence appropriate for the M32 field. }
\end{figure}

\begin{figure}
\includegraphics[width=90mm]{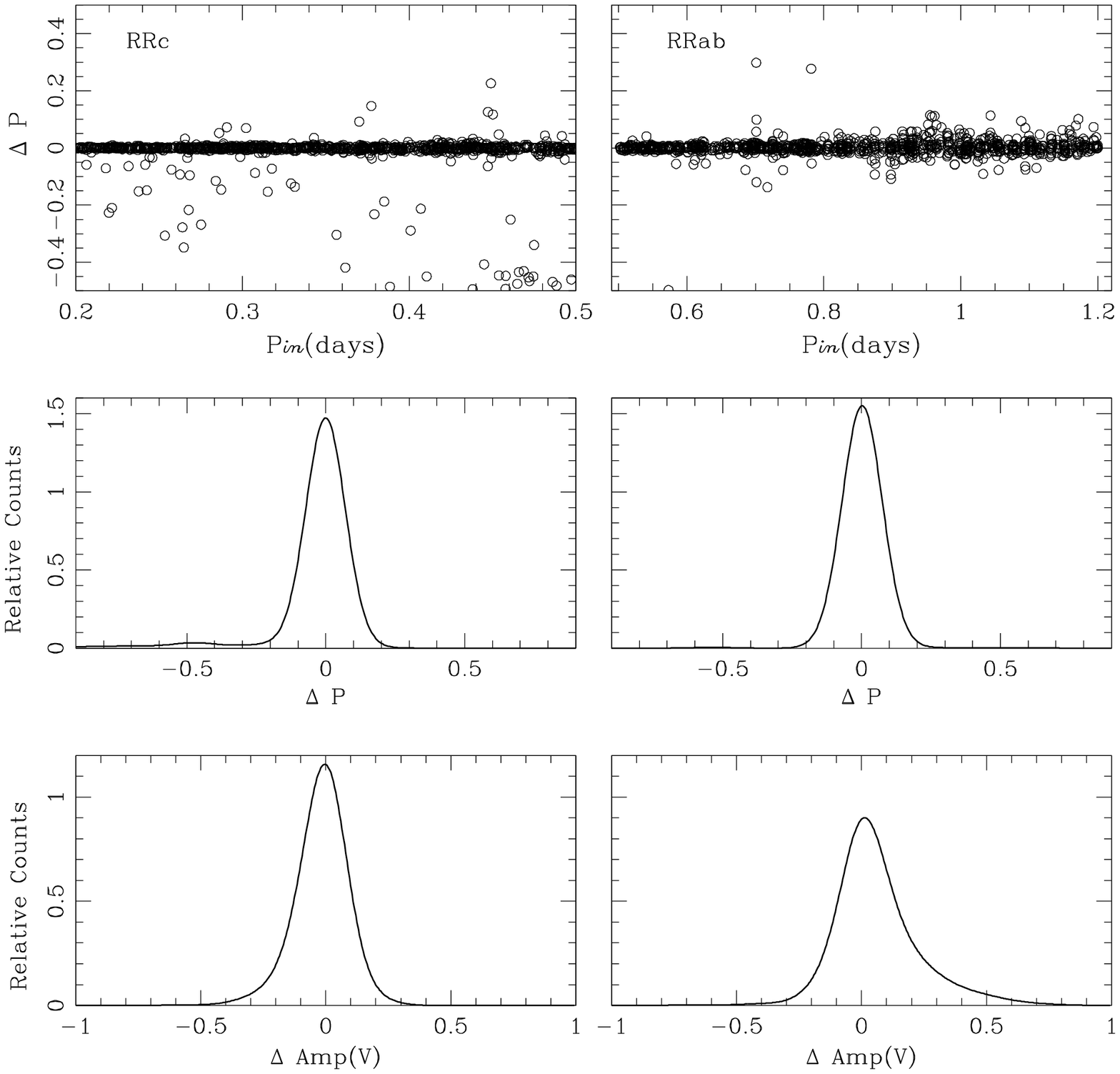}
\caption{The results of our RR Lyrae light curve simulations using an observing
cadence appropriate for the Control field. }
\end{figure}

\begin{figure}
\includegraphics[width=90mm]{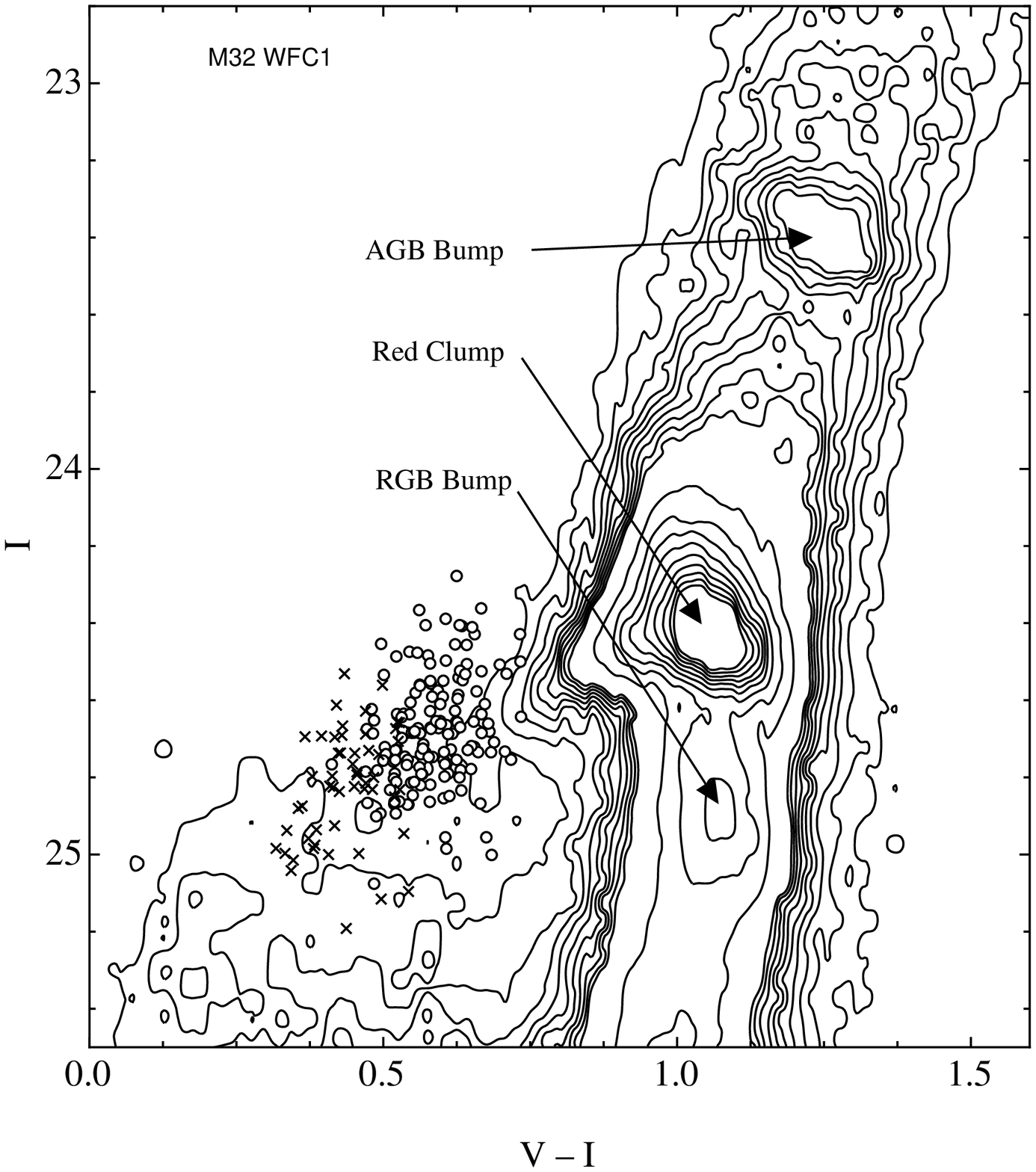}
\caption{The color-magnitude diagram (CMD) of stars measured on the WFC1 chip
of the M32 field. The ab-type
RR Lyraes are indicated as open circles while the c-types are shown as
crosses.}
\end{figure}

\begin{figure}
\includegraphics[width=90mm]{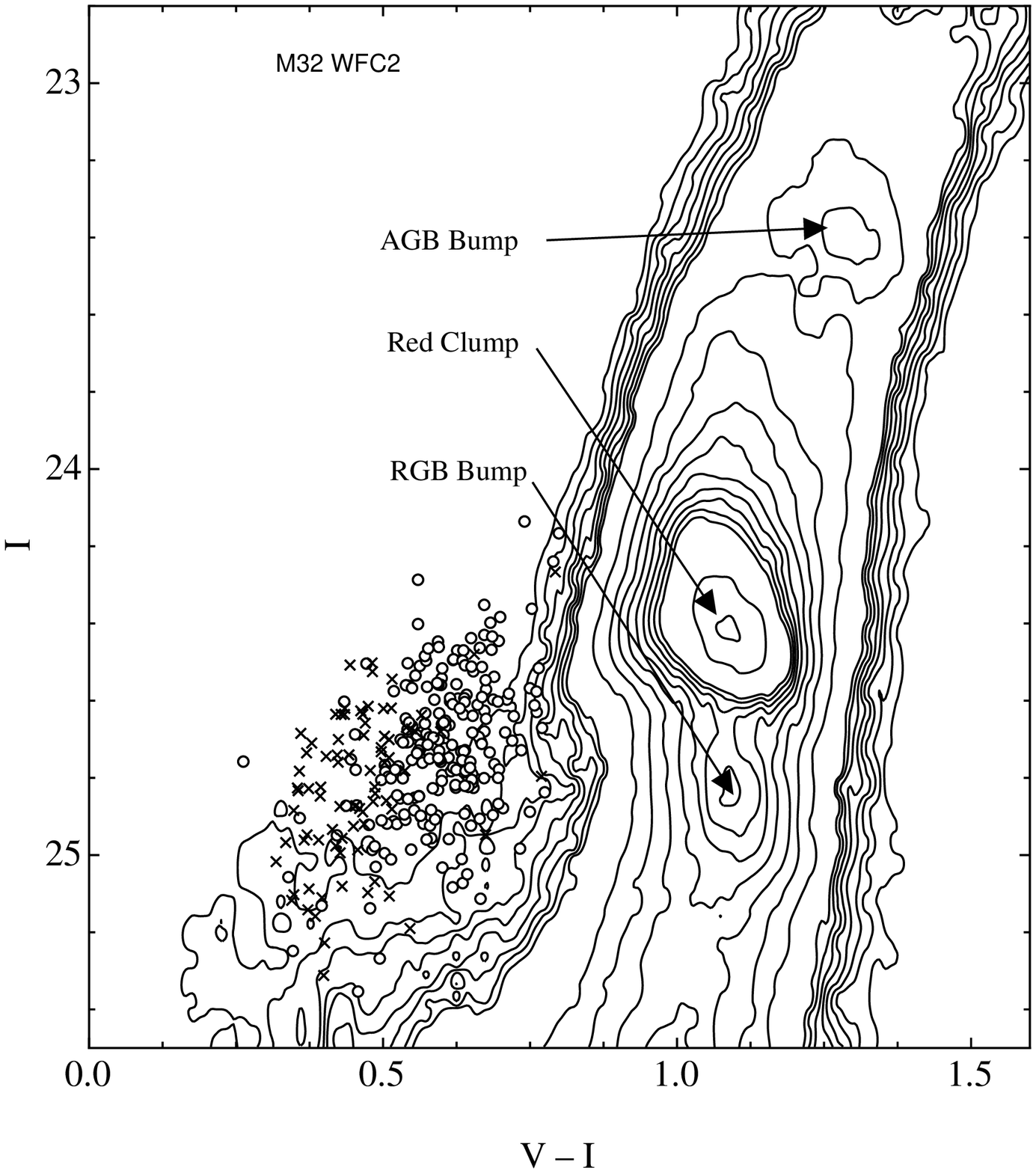}
\caption{The color-magnitude diagram (CMD) of stars measured on the WFC2 chip
of the M32 field. The ab-type
RR Lyraes are indicated as open circles while the c-types are shown as
crosses.}
\end{figure}

\section{RR Lyrae variables in M32}

We first examine the properties of the RR Lyrae variables in M32.
Figures 7 and 8 show the color-magnitude diagrams (CMDs) of the WFC1 and
WFC2 chips in the M32 field. Both CMDs show a strong red giant branch
(RGB), RC, asymptotic giant branch bump ($\approx1$ mag
above the RC), and RGB bump ($\approx0.5$ mag below the RC).  These
features were first detected in the CMD presented by Monachesi et
al.\ (2011), constructed from an ACS/HRC field which is entirely
overlapped by our observations.  Monachesi et al.\ (2011) showed that
the CMD locations of those features indicate that the bulk of the
stellar population in M32 at $\sim 2 \arcmin$ from its center is 8--10
Gyr old. The WFC2 CMD clearly shows more scatter, which is not
unexpected given the fact that it is located closer to the center of
M32. The RR Lyrae variables are also plotted in these CMDs and occupy
a region that is typical for these stars.

\subsection{The Radial Density Profile of RR Lyraes around M32}

We are interested in establishing the membership of our RR Lyrae
sample with regard to whether the majority of these stars belong to
M32 or the background M31 population. To do so, we show the radial
density distribution of these stars projected onto the major axis of
M32 as compared with the RC stars in Fig. 9. The red
clump stars are core-helium burning horizontal branch stars in the
same phase of evolution as the RR Lyraes. To isolate these stars, we
use a $I$-band magnitude range of 24.2--25.2 and a color range of
0.8--1.2. The upper panel of Fig. 9 illustrates the radial density
profiles of the RC and RR Lyrae stars from the present study where we have
applied photometric completeness corrections based on the artificial
star experiments described earlier.  We also include the RR Lyraes
from the F1 HRC field of Fiorentino et al.\ (2010, cross) and from
Field 1 of S09 (open squares). The profiles of the RR Lyrae stars have been
scaled to match the RC stars in the region outside of 150 arcsec. The
dashed line in the upper panel of Fig. 9 is our adopted M31
background RR Lyrae density of 0.0077 stars/sq arcsec as derived from
the outermost RR Lyrae point based on the observations in the present
study. We note that the average density of the seven outermost points
from the S09 study ($R_{major~axis}\geq 300\arcsec$) is $0.0068 \pm
0.00070$ stars/sq arcsec, which is consistent with our adopted value.

Adopting the outermost radial point as representative of the M31
background, we subtract this from the RC and RR Lyrae distributions and
plot the result in the lower panel of Fig. 9 along with the M32 major
axis surface brightness profile from Choi et al.\ (2002).  The latter
has been scaled to fit the RC profile outside of 90 arcsec. We see
that the RC and RR Lyrae star radial distributions agree from
$\approx100\arcsec$ out to the limit of the data. These in turn are
consistent with the shape of the surface brightness profile from Choi
et al.\ (2002). In contrast, the innermost RR Lyrae point at
$\approx77\arcsec$ is significantly below the RC profile at that
location. This is most likely a result of the fact that our detection
and characterization of RR Lyraes in the inner regions of M32 has been
adversely affected by the high stellar crowding. These are not
accounted for in our assessment of photometric completeness and are
more pronounced as the center of M32 is approached.

As mentioned above, Fig. 9 shows that the RR Lyraes follow the
RC radial profile very closely, which in turn follows the M32 surface
brightness profile from Choi et al.\ (2002). This suggests that a
significant fraction of the RR Lyraes in this ACS field belong to
M32. As a result, this is evidence that M32 contains RR Lyrae
variables and therefore a population older than $\approx10$ Gyr.

In order to estimate the number of RR Lyraes that likely belong to M32
in this field, we integrate the RR Lyrae radial profile in the lower
panel of Fig. 9, which is the one that is background-subtracted and
completeness corrected. Doing this, we find 222 RR Lyraes in this
field that belong to M32. This can be compared with the results of
Fiorentino et al.\ (2012), who claimed a total of 83 RR Lyraes
belonging to M32 in this field. The difference between these two
estimates likely stems from two sources. First, we have identified and
characterized 509 RR Lyraes in the M32 field compared with 416 from
Fiorentino et al.\ (2012) -- a factor of 1.22 more objects. Second,
the background density of M31 RR Lyraes in the present paper is
approximately a factor of 1.4 lower than that assumed by Fiorentino et
al.\ (2012) -- 0.0077 stars/sq arcsec compared with 0.011 stars/sq
arcsec. If we use the background density adopted by Fiorentino et
al.\ (2012), we arrive at a total number of 71 RR Lyraes belonging to
M32, which, when multiplied by a factor of 1.22, yields 87 RR Lyraes,
which is close to the value of 83 quoted by Fiorentino et al.\ (2012). As a
result, the main source of the difference between the present work and
that of Fiorentino et al.\ (2012) with regard to the number of RR
Lyraes belonging to M32 is the adopted contribution of such stars from
the M31 background.

\begin{figure}
\includegraphics[width=90mm]{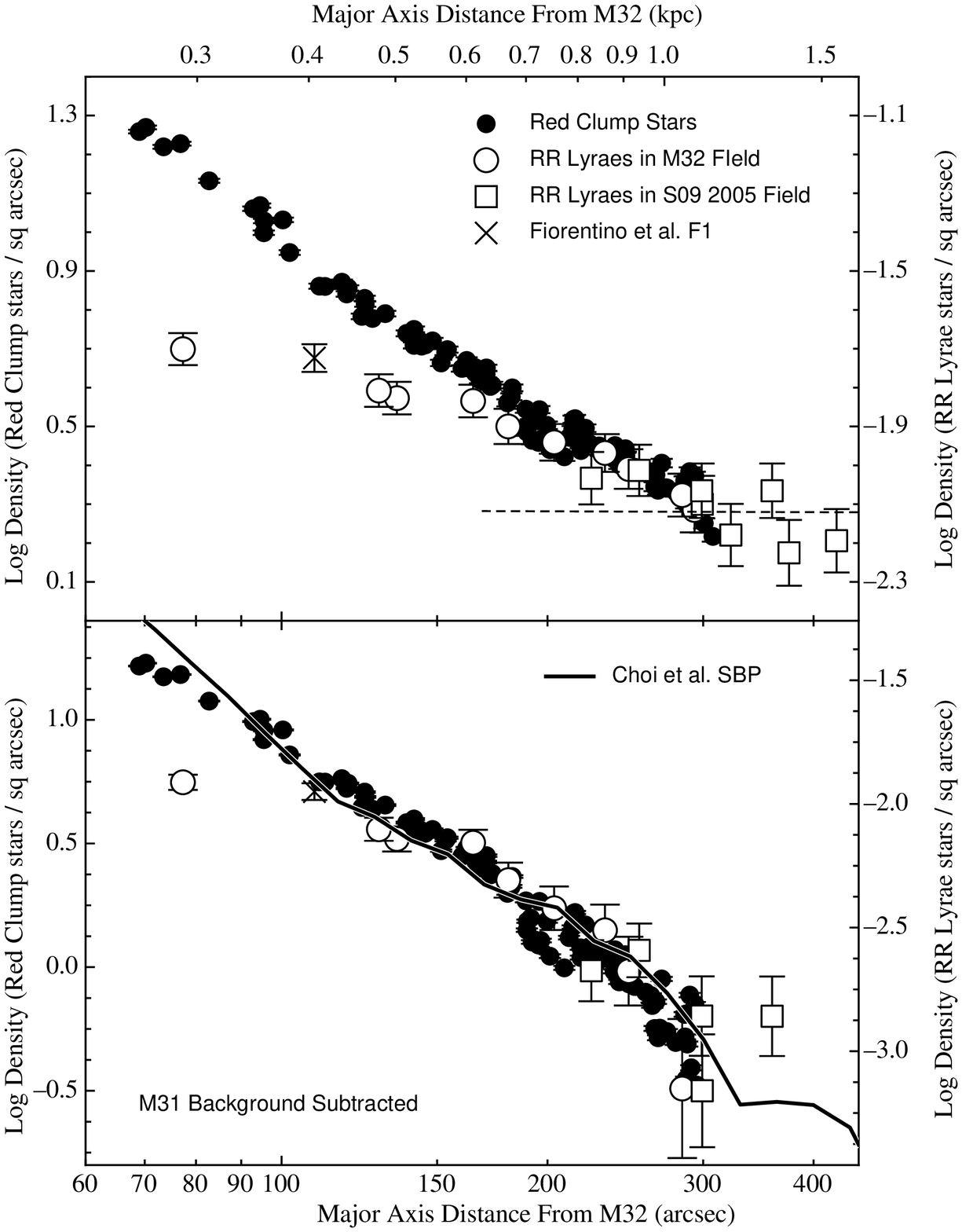}
\caption{The radial density profiles of red clump stars and RR Lyrae
  variables projected along the major axis of M32. The upper panel
  shows the raw profiles corrected for photometric incompleteness
  before the subtraction of the M31 background. In addition to the red
  clump stars (filled circles) and RR Lyraes (open circles) from the
  present study, we also include the RR Lyraes from the Fiorentino et
  al.  (2010, cross) F1 HRC field and from Field 1 of S09 (open
  squares). The profiles of these two populations have been scaled to
  match in the region outside of 150 arcsec. The upper abscissa
  assumes a M32 distance of 770 kpc. The dashed line is our adopted
  M31 background RR Lyrae level. In the lower panel, we have
  subtracted the M31 background density from the red clump stars and
  RR Lyraes and also include the M32 major axis surface brightness
  profile (SBP) from Choi et al.\ (2002) scaled to fit the red clump
  and RR Lyrae distribution. }
\end{figure}

\subsection{The Bailey Diagram of RR Lyrae Variables Associated with
  M32}

Figure 10 shows the Bailey Diagrams for the RR Lyraes in the M32
field. The ab-type RR Lyraes are shown by the filled points while the
c-type variables are plotted as open circles. The Oosterhoff I and II
loci are also indicated (Clement \& Rowe 1999). A comparison of these
diagrams with Fig.~10 of S09 suggests that the RR Lyraes in the M32
field are located slightly to the left (shorter period) of the
Oosterhoff I line. We note that this field has no significant
population of Oosterhoff II RR Lyraes.  We return to this point in
Section 6.1 below.

\begin{figure}
\includegraphics[width=90mm]{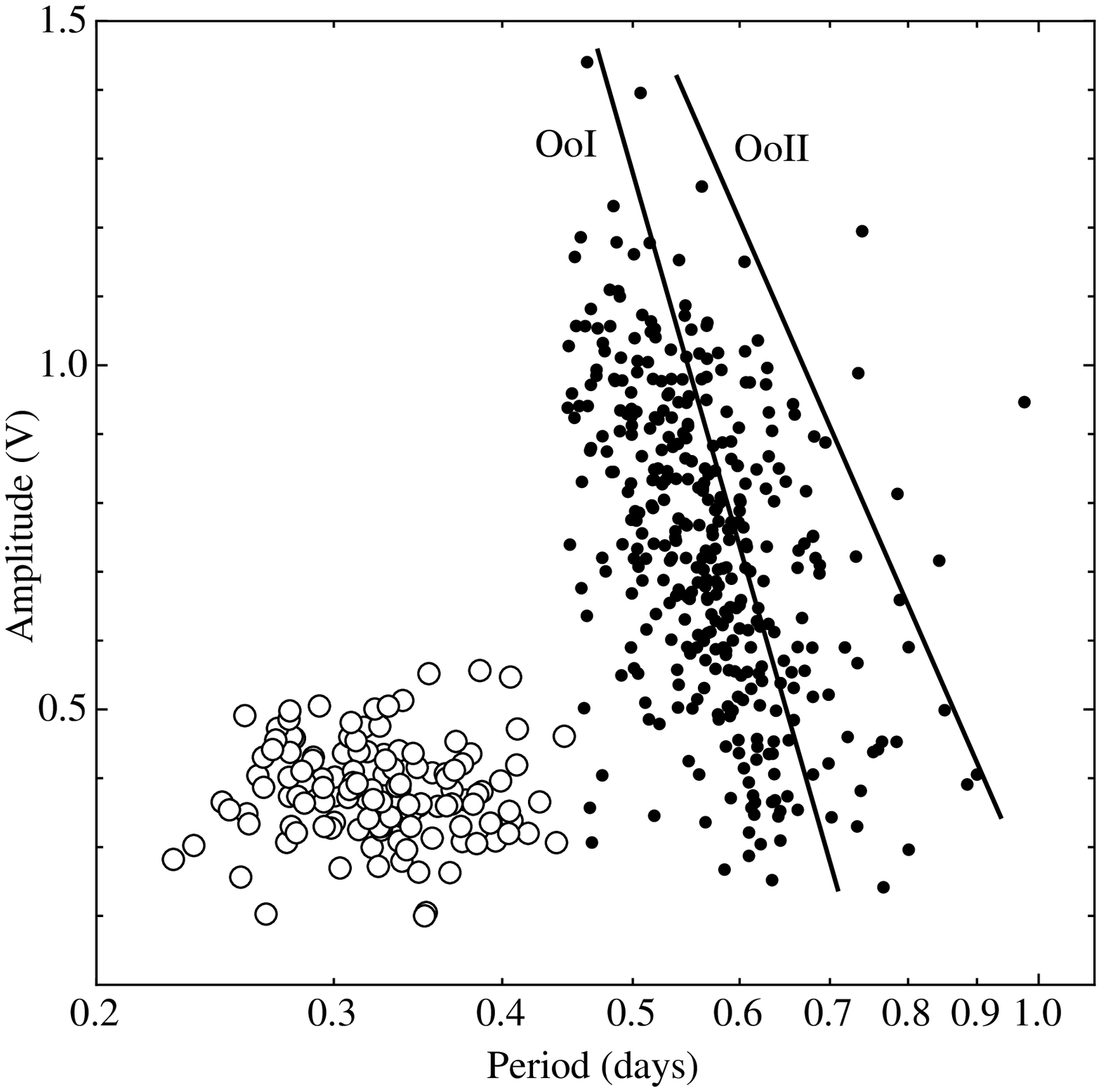}
\caption{The Bailey Diagram for the RR Lyraes in the M32 field showing
  V amplitude on the ordinate and period in days on the abscissa. The filled
  circles represent the ab-type RR Lyraes while the open circles
  are the c-types. The
  loci of ab-type RR Lyraes in Oosterhoff I and II Galactic globular
  clusters from Clement \& Rowe (1999) are also shown.}
\end{figure}

For the sake of completeness, we note that the mean periods of the RR
Lyraes in the M32 field are $\langle P_{ab}\rangle = 0.575 \pm 0.004$
(sem) d and $\langle P_{c} \rangle = 0.326 \pm 0.005$ (sem) d. These
are to be compared with mean periods of $\langle P_{ab}\rangle =
0.559$ d and $\langle P_{c}\rangle = 0.326$ d for the Oosterhoff I
Galactic globular clusters (Jeffery et al.\ 2011).

\subsection{The Metallicity of RR Lyrae Variables Associated with M32}

We can calculate the metallicities of ab-type RR Lyraes in two ways.
Using the data of Layden (2005, private communication) for 132
Galactic RR Lyraes in the solar neighborhood, Sarajedini et
al.\ (2006) established a relation between period and metal abundance
of the form
\begin{eqnarray}
\mathrm{[Fe/H]} = -3.43 - 7.82 \log P_{ab}.
\end{eqnarray}

This equation does not take into account the amplitudes of the RR
Lyraes even though, as the Bailey Diagrams show, there is a relation
between amplitude and period for the ab-types. Including this effect,
Alcock et al.\ (2000) gave a period-amplitude-metallicity relation of
the form
\begin{eqnarray}
\mathrm{[Fe/H]} = -8.85[\log P_{ab} + 0.15 A(V)] -2.60,
\end{eqnarray}
where $A(V)$ represents the amplitude in the $V$-band.  S09 suggest that if
the amplitudes of the RR Lyraes are well-determined, then equation (2)
yields more precise metal abundances than equation (1). As a result,
we will use equation (2) exclusively for the remainder of this paper.

For the 375 RRab stars in the M32 field, the solid line in Fig. 11
shows the distribution of [Fe/H] values given by equation (2). We find
an average metal abundance of $\langle\mathrm{[Fe/H]}\rangle = -1.42
\pm 0.02$ (sem) dex for these M32 RR Lyraes.  The dashed line in
Fig. 11 illustrates the metallicity distribution from Field 1 of S09,
which is dominated by M31 RR Lyraes, where the average turns out to be
$\langle\mathrm{[Fe/H]}\rangle = -1.46 \pm 0.03$ (sem)
dex. Furthermore, Fiorentino et al.\ (2010) find an average metallicity
of $\langle\mathrm{[Fe/H]}\rangle = -1.52 \pm 0.10$ dex in their F1
field, which is contained within our M32 field. These three average
values are statistically indistinguishable to within the errors. We
note also that we investigated the presence of a radial metallicity
gradient among the RR Lyraes in M32 but found no evidence supporting 
such a gradient.

\begin{figure}
\includegraphics[width=90mm]{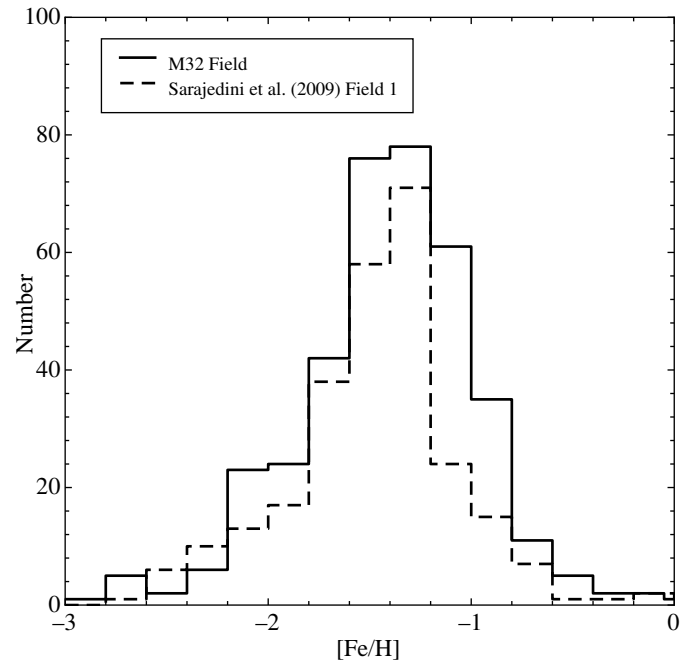}
\caption{The distribution of metallicities calculated using the
  equation from Alcock et al.\ (2000). The solid line represents
  ab-type RR Lyraes in our M32 Field while the dashed line is based on
  the field 1 RR Lyraes from S09.}
\end{figure}

\subsection{The Reddening and Distance to M32}

The work of Guldenshuh et al.\ (2005) showed that the minimum light
color of ab-type RR Lyrae variables is equal to $V-I=0.58\pm0.02$ mag
regardless of their metal abundances and pulsation properties such as
period and amplitude. Given this fact, we can calculate the reddening
for each RRab variable in our sample as shown in Fig. 12. The Gaussian
curve is fitted to the M32 Field RR Lyraes and shows that the shape of
the reddening histogram is consistent with a Gaussian distribution.
The error on any given reddening value is composed of $\pm0.02$ mag
from the Guldenshuh et al.\ (2005) calibration and $\pm0.05$ mag from
the uncertainty inherent in the measurement of the minimum light
colors. This latter value is taken from the results of our synthetic
light curve simulations described above. Taken together, these two
errors suggest that the majority of the dispersion in the reddening
distribution is attributable to errors inherent in the measurement of
the RR Lyrae minimum light color as opposed to internal reddening due
to M31 and/or M32. The mean reddening for the ab-type RR Lyraes in the
M32 field is $\langle E(V-I)\rangle = 0.134 \pm 0.088$ (sdm) $\pm
0.005$ (sem) mag.
We note that the Schlegel, Finkbeiner \& Davis (1998) dust maps
suggest a line-of-sight reddening to M32 of $E(B-V) = 0.08$ mag, which
translates to $E(V-I) = 0.11$ mag (Tammann, Sandage \& Reindl 2003).
This is consistent with our mean $E(V-I)$ values based on the minimum
light colors of the ab-type RR Lyraes thereby corroborating the
findings of previous investigators that M32 is largely free of
interstellar dust (Ford et al.\ 1978; Impey et al.\ 1986; Lauer et
al.\ 1998; Corbin et al.\ 2001; Choi et al.\ 2002).

\begin{figure}
\includegraphics[width=90mm]{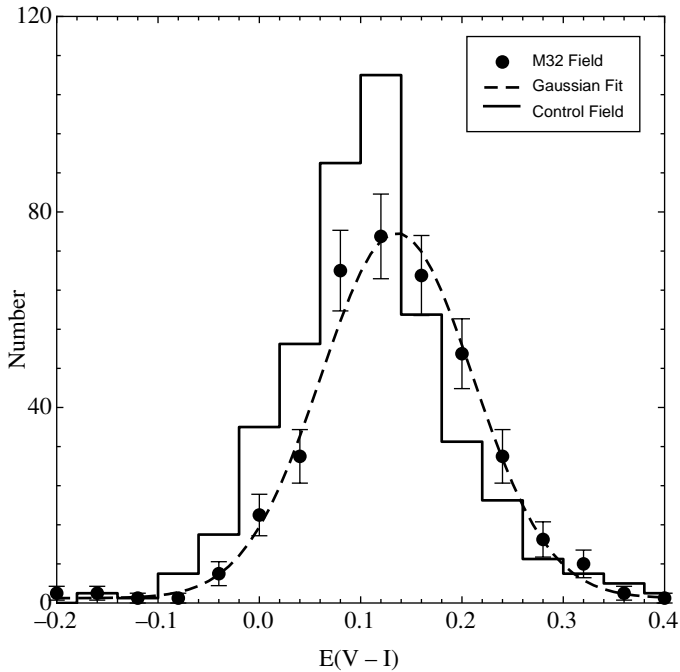}
\caption{The distribution of reddenings calculated from the ab-type RR
  Lyraes in the M32 and control fields. The dashed line is the
  Gaussian fit to the points.}
\end{figure}

With a determination of the line of sight reddening and the mean
metallicity of the ab-type RR Lyrae variables, we can now calculate
the distance to M32. We use the RR Lyrae luminosity--metallicity
relation of Chaboyer (1999): $M_V(RR) = 0.23\mathrm{[Fe/H]} + 0.93$
mag. Adopting $E(B-V) = 0.08 \pm 0.03$ mag and
$\langle\mathrm{[Fe/H]}\rangle = -1.42 \pm 0.20$ (zeropoint error in
the abundance scale) dex, along with $\langle V\mathrm{(RR)}\rangle =
25.28 \pm 0.05$ (zeropoint error from Sirianni et al.\ 2005) mag, we
find $\langle(V-M)_o\rangle = 24.42 \pm 0.12$ mag for the distance to
M32.  This distance is consistent with a number of previous authors:
$24.55 \pm 0.08$ mag (Tonry et al.\ 2001), $24.39 \pm 0.08$ mag (Jensen
et al.\ 2003), and $24.53 \pm 0.12$ mag (Monachesi et al.\ 2011).

\subsection{The Specific Frequency of M32 RR Lyraes and the Ancient
  Metal-Poor Population of M32}

We now compute the specific frequency of RR Lyrae stars in M32.  With
this number we can hope to understand the ancient, metal-poor
population in M32 in the context of Galactic globular clusters.
Suntzeff, Kinman \& Kraft (1991) define the specific frequency of RR
Lyrae stars $S_{RR}$ as the number of RR Lyrae stars $N_{RR}$ per unit
luminosity, normalized to a typical Galactic globular cluster
luminosity of $M_{Vt}=-7.5$ mag:
\begin{equation}
  S_{RR}=N_{RR} / 10^{-0.4(M_{Vt}+7.5)}
\end{equation}
(following the notation of Harris 1996).

In order to compute $S_{RR}$ for M32 RR Lyrae variables, we need to
determine the luminosity of old, metal-poor M32 stars contained in the
M32 ACS/WFC field.  We first compute the total $B$-band magnitude of
M32 in this field by integrated the $B$-band surface brightness
profile of Choi et al.\ (2002) within the ACS/WFC field; this gives
$B=11.85$ mag.  The average color of M32 beyond one effective radius
is $(B-V)=0.88$ mag (de Vaucouleurs et al.\ 1991), so $V=10.97$ mag.
Combined with the distance modulus and extinction (assuming
$A_V/E(B-V)=3.315$, Schlegel et al.\ 1998) determined in the previous
section, we find that the total $V$-band luminosity in the M32 field
is $M_{Vt}=-13.7$.  Monachesi et al. (2011) determined an upper limit of $\approx4.5$\% of
M32's mass was contained in stars with $[\mathrm{Fe/H}]<-1$, or a total
of $\approx5.5$\% of the $V$-band light from an analysis of the CMD of
the RGB; Monachesi et al. (2012) determined a lower limit of $\approx1$\% of M32's mass
was contained in similarly metal-poor stars from analysis of the CMD
near the (young) main-sequence turnoffs, or a total of $\approx1.3$\%
of the $V$-band light.  We use these two values as bounds on the
ancient, metal-poor starlight in the M32 ACS/WFC field; these
correspond to $M_{Vt}\approx-10.6$ mag (upper limit) and
$M_{Vt}\approx-9.0$ mag (lower limit) in this population.  Using these
luminosities, we finally derive $13 \la S_{RR} \la 56$ for M32's RR
Lyrae population (for the upper and lower limits to the mass of the
ancient, metal-poor stars in M32, respectively).  The lower limit is a
factor of two larger than the value found in the much smaller ACS/HRC
field by Fiorentino et al.\ (2010) due to our assumption of a smaller
fraction of metal-poor light than assumed in that paper (5.5\% versus
11\%; the larger fraction was based on a preliminary analysis of the
ACS/HRC data presented in Monachesi et al.\ 2011).  On the other hand,
the range of possible values falls within the range of Galactic
globular clusters: Brown et al.\ (2004) find that $1\la S_{RR}\la50$
at $[\mathrm{Fe/H}]\approx-1.4$ dex for Galactic globular clusters
using the data presented in Harris (1996).

As discussed in Fiorentino et al.\ (2010), the large dispersion in
$S_{RR}$ in Galactic globular clusters with $[\mathrm{Fe/H}]<-1$ dex
(due to the ``second-parameter problem'') means that there is no way
to invert this relation to determine the \emph{true} fraction of
ancient, metal-poor stars in a population; nor do there appear to be
stellar evolution models currently available that predict the
population of RR Lyrae \emph{a priori} (see, e.g., Salaris \& Cassisi
2005).  The presence of RR Lyrae variables tells us 
that ancient, metal-poor stars are
\emph{definitely} present in M32 -- a result that cannot be clearly
demonstrated by existing CMDs of M32 (e.g., Monachesi et al.\ 2011; 
Monachesi et al.\ 2012).  This in itself
is a very useful result, of course,
as it shows that at least some phase
of M32's evolution included metal-poor stars (Fiorentino et al.\ 2010, 2012).

\subsection{Comparison with Previous Studies}

The work of Fiorentino et al.\ (2010) imaged a field approximately 1.8
arcmin from the center of M32, which is wholly contained within the
M32 field analyzed herein. They found 17 RR Lyraes in their HRC field
of which 13 are recovered in our photometry. These numbers are not
unexpected given the fact that our photometric completeness at this
location is 88\% compared to 100\% for the Fiorentino et al.\ (2010)
study. We have matched the RR Lyraes in common and compared the
derived periods, amplitudes, and magnitudes. For the periods, there is
a mean difference of $\langle\Delta P\rangle = -0.0017 \pm 0.0048$
(sem) days in the sense (Fiorentino$-$Us). In the case of the RR Lyrae
amplitudes, we have adjusted for the difference in amplitudes expected
in the F555W (Fiorentino et al.\ 2010) and F606W filters (Present
Work) by decreasing the latter by 8\%, based on the work of Brown et
al.\ (2004) and S09. After making this adjustment, we calculate a
difference of $\langle\Delta\mathrm{Amp}(V)\rangle = 0.016 \pm 0.035$
(sem). The mean V magnitude difference is $\langle\Delta V\rangle =
0.016 \pm 0.029$ (sem), again in the sense (Fiorentino$-$Us). We see
that all of these differences are quite small and well within the
quoted errors.

In the case of the Fiorentino et al.\ (2012) RR Lyrae sample, they do
not provide tables of RR Lyrae data in their paper; however, we can
compare the mean properties of these stars between our study and
theirs. For example, Fiorentino et al.\ (2012) find $\langle
P_{ab}\rangle = 0.55 \pm 0.07$ (sdm) and $\langle P_{c}\rangle = 0.32
\pm 0.04$ (sdm). In the case of the present study, our analysis yields
$\langle P_{ab}\rangle = 0.575\pm 0.004$ (sem) and $\langle
P_{c}\rangle = 0.326 \pm 0.005$ (sem). The average periods of the
ab-type and c-type RR Lyraes are in good accord between the two
studies. In addition, Fiorentino et al.\ (2012) quote a mean intrinsic
$V$ magnitude of $\langle V_{o}\rangle = 24.95 \pm 0.18$ mag. Given
their adopted reddening of $E(B-V) = 0.08$, this translates to a mean
apparent magnitude of $\langle V(\mathrm{RR})\rangle = 25.20$, which is in
agreement with our value of $\langle V(\mathrm{RR})\rangle = 25.30 \pm
0.05$.

\section{RR Lyrae Variables in M31}

We now turn to the properties of RR Lyrae variables in M31,
concentrating on the Control field and its implications for the M31
spheroid.

\subsection{The Bailey Diagram of M31 RR Lyrae Variables}

Figure 13 shows the Bailey Diagram for the RR Lyraes in the Control
field (see Fig.~10 for the M32 population). In contrast to the M32
field, the RR Lyraes in the Control field reveal an intriguing bimodal
appearance with two distinct ab-type sequences roughly bisected by the
Oosterhoff type I locus. This is the first time that such bimodal
behavior has been observed among the field RR Lyrae population of
M31. We investigate the significance of this phenomenon in the
following sections. For now, we note, as for M32, this field does not
have a significant population of Oosterhoff II RR Lyraes. As noted by
S09 and Yang \& Sarajedini (2012), this further underscores the
general dearth of Oosterhoff II RR Lyraes in the spheroid of M31.  In
fact, almost all of the M31 fields observed so far -- except for the
Control field considered here -- and M32 exhibit Bailey Diagrams that
resemble each other, i.e., they are all aligned with or slightly
offset to shorter periods compared with the Oo I line.
For completeness, we find $\langle P_{ab}\rangle = 0.568 \pm 0.004$
(sem) d and $\langle P_{c} \rangle = 0.333 \pm 0.003$ (sem) d for the
Control field.

In order to investigate the robustness of the bimodality in Fig. 13, we
take each ab-type RR Lyrae in the Bailey Diagram and calculate its period 
difference relative to the Oosterhoff I line shown in Fig. 13. Since the 
Oosterhoff I locus roughly bisects the two subpopulations as noted above, 
it is a natural fiducial for this purpose. The resultant histogram of period 
differences is shown as the filled points in Fig. 14 wherein the solid line histogram 
is the Field 2 data from Sarajedini et al. (2009). The latter distribution has been 
shifted horizontally by --0.03d in order to match the dominant peaks of the two 
histograms so that they can be compared in a relative manner. We see that 
the secondary peak present in the Control field histogram is not present in field 2 
of Sarajedini et al. (2009). It is also possible to compare the control field 
histogram shown in Fig. 14 with a single-peaked Gaussian distribution with 
the same mean and standard deviation via the Kolmogorov-Smirnov (K-S) test. 
We find that there is only a 1\% chance that the two distributions are 
from the same parent population. This suggests that the bimodal appearance
of the Control field Bailey Diagram is likely to be genuine.

\begin{figure}
\includegraphics[width=90mm]{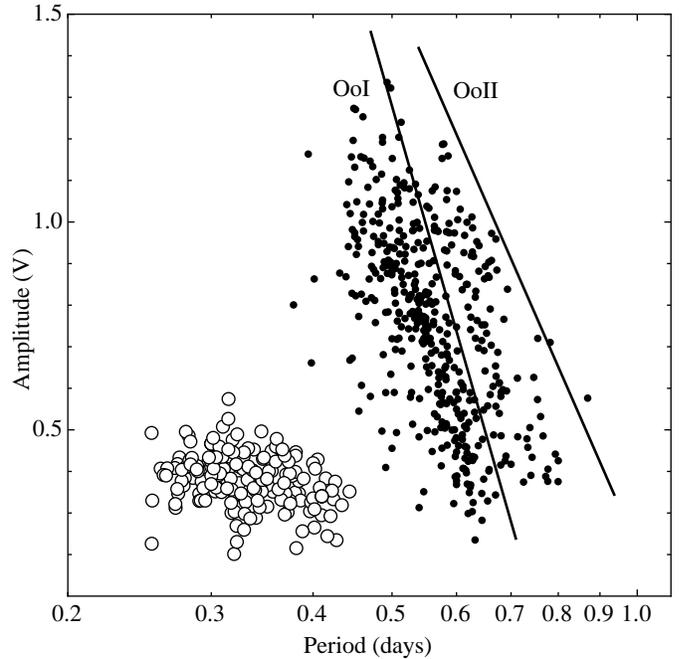}
\caption{The Bailey Diagram for the RR Lyraes in the Control field
  showing V amplitude on the ordinate and period in days on the
  abscissa. The filled
  circles represent the ab-type RR Lyraes while the open circles
  are the c-types. The loci of ab-type RR Lyraes in Oosterhoff I and II
  Galactic globular clusters from Clement \& Rowe (1999) are also
  shown.}
\end{figure}

\begin{figure}
\includegraphics[width=90mm]{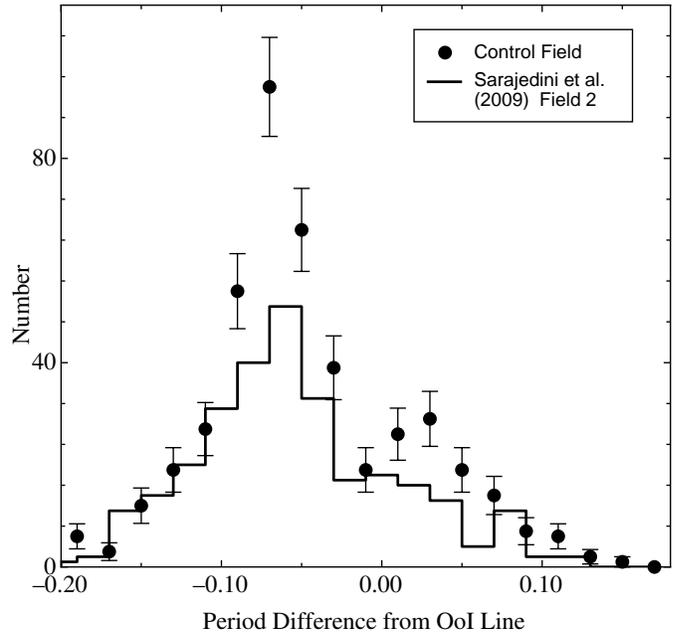}
\caption{The filled circles represent the histogram of period differences for 
each ab-type RR Lyrae in the Control field relative to the Oosterhoff I line 
shown in Fig. 13. The solid line is the histogram of period differences
measured in the same way for the ab-type RR Lyraes in Field 2 of Sarajedini
et al. (2009). The latter has been shifted horizontally by -0.03d in order
to match the dominant peaks of the two histograms.}
\end{figure}

\subsection{The Metallicities of M31 RR Lyrae Variables}

Following the analysis of Section 5.3 above, the metallicity
distribution function of the 446 ab-type RR Lyraes in the Control
field is shown in Fig. 15. Not surprisingly, given the bimodal
appearance of the Bailey Diagram, these RR Lyraes exhibit a bimodal
metallicity distribution with a well-pronounced peak at
$[\mathrm{Fe/H}]\approx-1.3$ dex and another at
$[\mathrm{Fe/H}]\approx-1.9$ dex.  This is in contrast to the single
peaked distributions of the M32 field and all previous M31 fields
published to date (e.g. S09; Fiorentino et al.\ 2010; Jeffery et
al.\ 2011). Figure 15 compares the Control field MDF with that of the
RRab stars in field 2 from S09, which has a mean abundance of
$\langle\mathrm{[Fe/H]}\rangle = -1.54 \pm 0.03$ (sem) dex.  Upon
closer examination, it would appear that the latter may exhibit a
metal-poor tail that corresponds to the metal-poor peak shown by the
Control field. In any case, the bimodal appearance of Figs. 14 and 15 
suggest that the Control field line-of-sight intersects two old stellar
populations with distinctly different mean metallicities. It is
possible that the primary peak represents a stellar population
belonging to the putative M31 spheroid while the more metal-poor
population could have originated in a disrupted M31 dwarf satellite
galaxy. If so, then the bimodal appearance of Fig. 13 represents more
evidence supporting the merger and accretion history of the M31
spheroid (McConnachie et al.\ 2009).

\begin{figure}
\includegraphics[width=90mm]{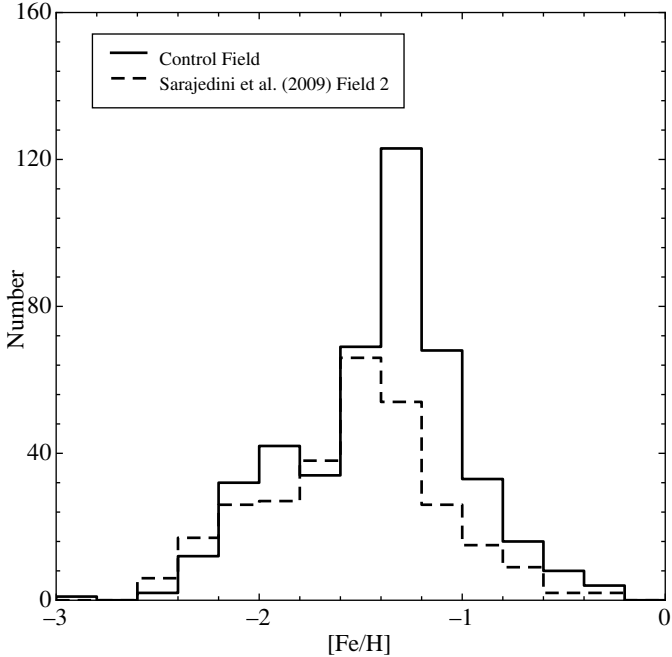}
\caption{The distribution of metallicities calculated using the
  equation from Alcock et al.\ (2000). The solid line represents
  ab-type RR Lyraes in our Control Field while the dashed line is
  based on the field 2 RR Lyraes from S09.}
\end{figure}

\subsection{The Reddening and Distance to M31}

Following the discussion in Section 5.4, we find for the Control field
an average reddening of $\langle E(V-I)\rangle = 0.111 \pm 0.088$
(sdm) $\pm 0.004$ (sem) mag (see Fig. 12).  Using this value and a mean metallicity
of $\langle\mathrm{[Fe/H]}\rangle = -1.39 \pm 0.20$ dex for the
Control field, we find a mean magnitude of the RR Lyraes in this field
of $\langle V(\mathrm{RR})\rangle = 25.27 \pm 0.05$ mag. Once again
adopting $E(B-V) = 0.08 \pm 0.03$, we infer a distance modulus of
$\langle (V-M)_o\rangle = 24.41 \pm 0.12$ mag, which is in agreement
with the result from the M32 field. We can also proceed slightly
differently here by dividing the ab-type RR Lyraes in the Control
field at $\mathrm{[Fe/H]} = -1.7$ dex and calculating the distance for
each group separately. We find that those with $\mathrm{[Fe/H]}<-1.7$
dex exhibit a distance modulus of $\langle (V-M)_o\rangle = 24.43\pm
0.12$ mag while those with $\mathrm{[Fe/H]}\geq-1.7$ dex have
$\langle(V-M)_o\rangle = 24.41 \pm 0.12$ mag. These are essentially
the same indicating that, to the level of precision with which we are
able to measure it, the two stellar population components in the
Control field are at the same distance. Furthermore, our distances are
consistent with a number of previous values for the true distance
modulus of M31, e.g. $24.44 \pm 0.11$ mag (Freedman \& Madore 1990),
$24.50 \pm 0.10$ mag (Brown et al.\ 2004), and $24.47 \pm 0.07$ mag
(McConnachie et al.\ 2005).



\subsection{RR Lyraes in the M31 Spheroid}

We can gain insight into the broader properties of RR Lyraes in M31 by
examining the radial distribution of the RR Lyrae populations thus far
studied. Figure 16 illustrates these data wherein the top panel shows the
radial density distribution of M31 RR Lyraes compared with three M31 minor
axis surface brightness profiles. 
The two inner data points come from S09 where `halo4' represents
their Field 2 and `halo6' is their Field 1. The points designated
`M32' and `Control' are from the present work.  The halo11, halo21,
halo35a, and halo35b fields are the minor axis fields taken from the
work of Jeffery et al.\ (2011), while the stream and disk points also
come from Jeffery et al.\ (2011). The warp and outer disk fields are
from the paper by Bernard et al.\ (2012).  Next to each point, we
indicate the number of RR Lyraes at that location. All of these data
are based on studies that have used HST/ACS/WFC in order to identify
and characterize the RR Lyrae variables. The solid line is the minor
axis surface brightness profile from Pritchet \& van den Bergh (1994,
see also Brown et al. 2008) scaled to match the inner two points from
S09 and the outer halo points (halo21, halo35a, halo35b). Also plotted 
are the minor axis
surface brightness profiles from Irwin et al. (2005, thin solid line) and
Gilbert et al. (2009, filled squares). We see that these three profiles agree
well with each other. In order to
be consistent with the minor axis surface brightness profiles, the
radial positions of the inner three M31 RR Lyrae fields have been projected onto
the minor axis of M31 and are plotted as such.  The halo11, halo21,
halo35a, and halo35b points are already along the M31 minor axis.

\begin{figure}
\includegraphics[width=90mm]{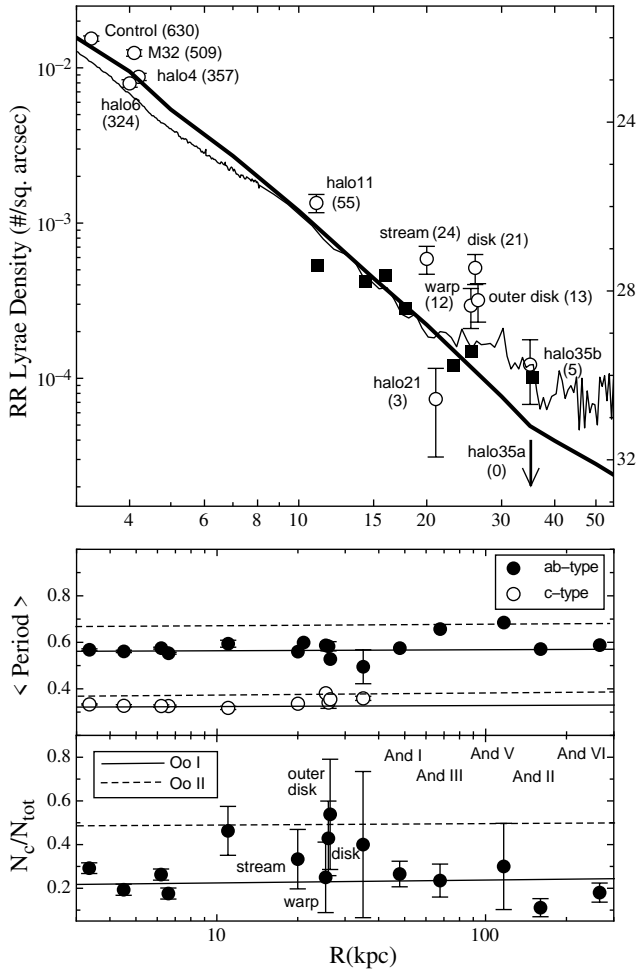}
\caption{The upper panel shows the radial density profile of RR Lyraes
  in the spheroid of M31. These are referenced to the left ordinate
  scale.  The solid line is the minor axis surface brightness profile
  from Pritchet \& van den Bergh (1994, see also Brown et al.\ 2008
  and Courteau et al. 2011)
  referenced to the right ordinate axis. Also plotted are the minor axis
  radial brightness profiles from Irwin et al. (2005, thin solid line) and
  Gilbert et al. (2009, filled squares). The middle panel displays the
  radial trend of the mean period for the ab-type (solid circles) and
  c-type (open circle) RR Lyrae variables.  The bottom panel is
  similar to the middle one except that the ratio of c-type to all RR
  Lyraes is illustrated as a function of radial location in the M31
  spheroid. The solid lines illustrate the properties of RR Lyraes in
  Oosterhoff I clusters while the dashed lines are for Oosterhoff II
  clusters. The properties of the Andromeda dwarf galaxies are taken
  from Table 5 of Clementini (2010). See text for more details.}
\end{figure}

\begin{figure}
\includegraphics[width=90mm]{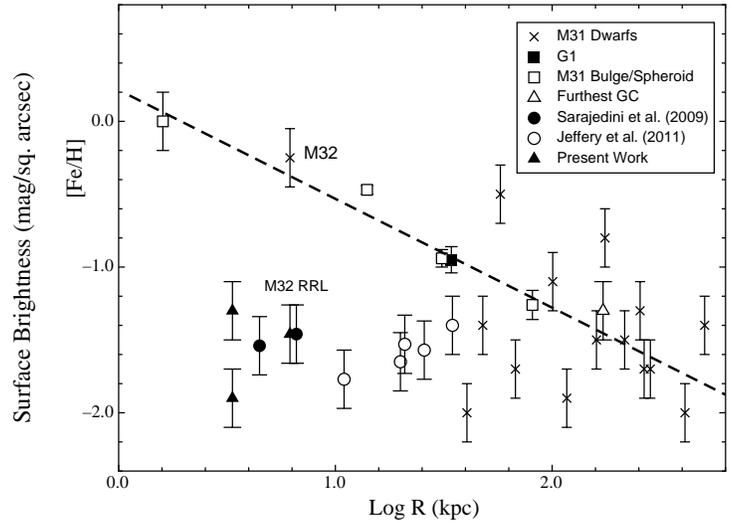}
\caption{The radial variation of metal abundance in the M31 spheroid
  as originally presented by Sarajedini et al. (2009, filled circles)
  augmented by the results of the present work (filled triangles) and
  those of Jeffery et al.\ (2011, open circles). The innermost open
  square represents the bulge abundance measured by Sarajedini \&
  Jablonka (2005). The remaining open squares are the bulge/halo
  points from the work of Kalirai et al.\ (2006). The dashed line is
  the least squares fit to these data with a slope of
  $-0.75\pm0.11$. The crosses represent the dwarf galaxies surrounding
  M31 from the work of Grebel et al.\ (2003) and Koch \& Grebel (2006)
  whereas the abundance of M32 is taken from Grillmair et
  al. (1996). The filled square is the well-known massive globular
  cluster G1 studied by Meylan et al.\ (2001). The open triangle is
  the furthest known globular cluster in M31 discovered by Martin et
  al.\ (2006). All of these points have been scaled to an M31 distance
  of $(m-M)_0 = 24.43$.}
\end{figure}

From the top panel of Fig. 16, we note that, not surprisingly, the
surface density of RR Lyrae variables in M31 drops with distance from
its center. The minor axis points (halo4, halo6, halo11, halo21,
halo35a, halo35b) more or less follow the plotted surface brightness
profile to within the errors. The stream,  disk, outer disk, and warp fields 
fall above the halo distribution suggesting that they are enhanced in 
RR Lyraes above what would be expected from the halo alone. This is 
consistent with the conclusions of Jeffery et al. (2011) and Bernard et al. 
(2012) that these RR Lyraes are genuine members of these subsystems. 
If this is the case, then the stream, disk, outer disk, and warp of M31 harbor an 
old stellar population suggesting that the earliest epoch of star formation in 
all of these subsystems occurred at approximately the same time.


The middle panel of Fig. 16 shows the variation of mean RR Lyrae
period with distance from the center of M31 for the fields shown in
the top panel as well as the data for five of the dwarf spheroidal
companions of M31 taken from Clementini (2010). The mean periods of
Oosterhoff I and II Galactic globular clusters are also shown as solid
and dashed lines, respectively, for comparison.  The upper set of
solid/dashed lines refer to ab-type variables while the lower set are
for c-type variables. These lines come from the work of Jeffery et
al.\ (2011).  We see in this diagram, that, with the exception of And
III and And V, the rest of the M31 halo is solidly of Oosterhoff I
type in terms of the mean periods of the RR Lyrae variables.

The bottom panel of Fig. 16 is similar to the middle panel except that
the ratio of c-type to all RR Lyraes is plotted as a function of
galactocentric distance. Once again, the M31 fields in the upper panel
are included as well as the dwarf satellites from Clementini
(2010). The conclusion we reached from the middle panel is also
largely unchanged -- to within the error bars, the M31 halo is best
described as containing RR Lyraes that are most similar to Oosterhoff
I Galactic globular clusters.

The reason that Oosterhoff I RR Lyraes dominate the M31 halo is
related to the variation of metal abundance with radial position as
discussed in S09. Figure 17 shows an updated version of Fig. 14 from
S09.  The metallicities for the RR Lyraes in the two fields considered
herein are shown by the filled triangles while the open circles
represent the RR Lyraes considered in the study of Jeffery et
al.\ (2011). The inner-most point is the bulge metallicity from the
work of Sarajedini \& Jablonka (2005), while the remaining open
squares are the bulge/halo points from the work of Kalirai et
al. (2006) as shown in their Table 3. The dashed line is the least
squares fit to the open squares with a slope of $-0.75 \pm 011$.  The
other points represent the dwarf spheroidal companions to M31
(crosses, Grebel et al.\ 2003; Koch \& Grebel 2006), the globular
cluster G1 (filled square, Meylan et al.\ 2001), and the furthest
globular cluster in M31 (open triangle, Martin et al.\ 2006). Note
that we have adopted the mean metallicity of M32 from the work of
Grillmair et al.\ (1996). All of the non-RR Lyrae metallicities are
based on the colors of RGB stars save for those from the study of 
Grillmair et al. (1996), which utilize integrated spectral indices.
All of these values are based on a distance
of $(m-M)_o = 24.43$ (770 kpc) for M31. In cases where an error in the
metallicity is not available, we have adopted a value of $\pm0.2$ dex.

Metal abundance is the primary parameter affecting horizontal branch
(HB) morphology among the Galactic globular clusters; this is also
true in terms of the Oosterhoff dichotomy. As shown in Table 2 of
Clementini (2010), Oosterhoff I clusters have
$\mathrm{[Fe/H]}\approx-1.4$ while those of Oosterhoff II have
$\mathrm{[Fe/H]}\approx-2.0$. As Fig. 17 reveals, the halo of M31
($R_{gc}\gea50\,\mathrm{kpc}$) has a mean metallicity that is closer
to $\mathrm{[Fe/H]}\approx-1.4$ than $\mathrm{[Fe/H]}\approx-2.0$,
thus corroborating its Oosterhoff I nature.

\section{Summary and Conclusions}

We have presented HST/ACS observations of two fields - one that 
samples the stellar populations of M32 and another dubbed `Control' that is 
dominated by the spheroid of M31.
Our photometry in these fields has facilitated the identification and characterization
of 1139 RR Lyrae variables of which 821 are ab-type and 318 are c-type.
Based on an analysis of the periods, amplitudes, magnitudes, and spatial
distributions of these stars, we draw the following conclusions.

\vskip 12pt
\noindent 1. We find a radial gradient in the density of RR Lyraes relative to the 
center of M32. This gradient is consistent with the surface brightness profile of M32 
suggesting that a significant number of the RR Lyraes in this region belong to M32. 
This provides further confirmation that M32 contains an ancient stellar population 
formed around the same time as the oldest population in M31 and the Milky Way. 

\vskip 12pt
\noindent 2. The ab-type RR Lyraes in M32 are closer to the Oosterhoff I line in the 
Bailey Diagram as compared to the Oosterhoff II locus exhibiting a mean metal abundance 
of $\langle\mathrm{[Fe/H]}\rangle = -1.42 \pm 0.02$ (sem) dex. 

\vskip 12pt
\noindent 3. The mean reddening we measure for the M32 RR Lyraes is consistent
with being due entirely to extinction within the Milky Way reinforcing the finding of
previous investigators that M32 contains little or no dust. Adopting a
mean reddening of $E(B-V) = 0.08 \pm 0.03$ mag, and a relation between
RR Lyrae luminosity and metallicity, we find an absolute distance modulus of
$\langle(V-M)_o\rangle = 24.42 \pm 0.12$ mag for M32 and 
$\langle (V-M)_o\rangle = 24.41 \pm 0.12$ mag for M31 based on the Control
field RR Lyrae variables.

\vskip 12pt
\noindent 4. In the Control field, the Bailey Diagram 
shows the unprecedented signature of two sequences among the ab-type
RR Lyraes. When interpreted in terms of metal abundance, the primary sequence
corresponds to a population of RR Lyraes with $\mathrm{[Fe/H]}\approx-1.3$ dex while
the secondary peak occurs at $\mathrm{[Fe/H]}\approx-1.9$ dex. 
We speculate that the primary peak represents the 
putative M31 spheroid while the more metal-poor population could have originated in a 
disrupted M31 dwarf satellite galaxy. 

\vskip 12pt
\noindent 5. An examination of the global properties of RR Lyraes in the
Andromeda system reveals that, with few exceptions, M31 and its satellite galaxies
contain Oosterhoff type I RR Lyraes. This is likely due to the fact that
the mean metal abundance of the M31 halo is more representative of Oosterhoff I
Galactic globular clusters than of Oosterhoff type II clusters. Needless to say,
the mere fact that all of these systems contain RR Lyraes suggests that their 
earliest epoch of star formation occurred at approximately the same time.

\section*{Acknowledgments}
This work has made use of the IAC-STAR Synthetic CMD computation code. IAC-STAR is 
supported and maintained by the computer division of the Instituto de Astrofisica de 
Canarias. We are grateful to the anonymous referee whose comments improved the clarity and quality of this manuscript. We also thank St\'ephane Courteau for
making available the M31 surface brightness profiles from Courteau et al.
(2011). We acknowledge support from NASA through grant AR-12153.01-A from the Space 
Telescope Science Institute, which is operated by the Association of Universities for 
Research in Astronomy, Inc., for NASA under contract NAS5-26555.

\bsp

\label{lastpage}


\begin{thebibliography}{99}
\bibitem[\protect\citeauthoryear{Alcock et al.}{2000}]{alcock2000} Alcock, C. et al. 2000, AJ, 119, 2194
\bibitem[\protect\citeauthoryear{Alonso-Garcia et al.}{2004}]{alonzo2004} Alonso-Garc\'{i}a, J., Mateo, 
M., \& Worthey, G. 2004, AJ, 127, 868
\bibitem[\protect\citeauthoryear{Aparicio \& Gallart}{2004}]{aparicio2004} Aparicio, A., \& Gallart, C.
2004, AJ, 128, 1465
\bibitem[\protect\citeauthoryear{Bernard et al.}{2012}]{bernard2012} Bernard, E. J., Ferguson, A. M. N.,
Barker, M. K., Hidalgo, S. L., Ibata, R. A., Irwin, M. J., Lewis, G. F., McConnachie, A. W.,
Monelli, M., \& Chapman, S. C. 2012, MNRAS, in press
\bibitem[\protect\citeauthoryear{Brown et al.}{2004}]{brown2004} Brown, T. M., Ferguson, H. C., Smith, 
E., Kimble, R. A., Sweigart, A. V., Renzini, A., \& Rich, R. M. 2004, AJ, 127, 2738
\bibitem[\protect\citeauthoryear{Brown et al.}{2008}]{brown2008} Brown, T. M. et al. ApJ, 685, L121
\bibitem[\protect\citeauthoryear{Chaboyer}{1999}]{chab1999}  Chaboyer, B. 1999, in Post-Hipparcos 
Cosmic Candles, ASSL, Vol. 237, edited by A. Heck and F. Caputo 
(Dordrecht: Kluwer Academic Publishers) p.111
\bibitem[\protect\citeauthoryear{Choi et al.}{2002}]{choi2002} Choi, P. I., Guhathakurta, P.,
\& Johnston, K. V. 2002, AJ, 124, 310
\bibitem[\protect\citeauthoryear{Clementini}{2010}]{clementini2010} Clementini, G. 2010, in Variable
Stars, The Galactic Halo, and Galaxy Formation, edited by C. Sterken, N. Samus, and L. Szabados, 
(Sternberg Astronomical Institute: Moscow State University) p. 107
\bibitem[\protect\citeauthoryear{Clement \& Rowe}{1999}]{clement1999} Clement, C. M. \& Rowe, J. 
2000, AJ, 120, 2579
\bibitem[\protect\citeauthoryear{Coelho et al.}{2005}]{coelho2005} Coelho, P., Mendes de Oliveira, 
C., \& Cid Fernandes, R. 2009, MNRAS, 396, 624
\bibitem[\protect\citeauthoryear{Corbin et al.}{2001}]{corbin2001} Corbin, M. R., O'Neil, E., \& Rieke, M. J. 2001, AJ, 121, 2549
\bibitem[\protect\citeauthoryear{Courteau et al.}{2011}]{courteau2011} Courteau, S.
Widrow, L. M., McDonald, M., Guhathakurta, P., Gilbert, K. M., 
Zhu, Y., Beaton, R. L., \& Majewski, S. R. 2011, ApJ, 739, 20
\bibitem[\protect\citeauthoryear{de Vaucouleurs et 
al.}{1991}]{1991rc3..book.....D} de Vaucouleurs G., de Vaucouleurs A., 
Corwin H.~G., Jr., Buta R.~J., Paturel G., Fouqu{\'e} P., 1991, 
Third Reference Catalogue of Bright Galaxies (New York: Springer)
\bibitem[\protect\citeauthoryear{Fiorentino et al.}{2011}]{fior2011} Fiorentino, G. Monachesi, A.,
Trager, S. C., Lauer, T. R., Saha, A., Mighell, K. J., Freedman, W., Dressler, A., Grillmair, C.,
\& Tolstoy, E. 2010, ApJ, 708, 817
\bibitem[\protect\citeauthoryear{Fiorentino et al.}{2012}]{fior2012} Fiorentino, G., Contreras 
Ramos, R., Tolstoy, E., Clementini, G., \& Saha, A. 2012, A\&A, in press
\bibitem[\protect\citeauthoryear{Ford, Jacoby, \& Jenner}{1978}]{1978ApJ...223...94F} Ford H.~C., Jacoby G.~H., Jenner D.~C., 1978, ApJ, 223, 94 
\bibitem[\protect\citeauthoryear{Freedman}{1992}]{freed1992} Freedman, W. L. 1992, AJ, 104, 1349
\bibitem[\protect\citeauthoryear{Freedman \& Madore}{1990}]{freedman1990} Freedman, W. L., \& Madore, B. F. 1990, ApJ, 365, 186
\bibitem[\protect\citeauthoryear{Gilbert et al.}{2009}]{gilbert2009} Gilbert, K. M., 
Font, A. S., Johnston, K. V., \& Guhathakurta, P. 2009, ApJ, 701, 776
\bibitem[\protect\citeauthoryear{Grebel et al.}{2003}]{grebel2003} Grebel, E. K., Gallagher, J. S. III, \& Harbeck, D. 2003, AJ, 125, 1926
\bibitem[\protect\citeauthoryear{Grillmair et al.}{1996}]{grill1996} Grillmair, C. J., et al. 1996, AJ, 112, 1975
\bibitem[\protect\citeauthoryear{Guldenschuh et al.}{2005}]{gulden2005} Guldenschuh, K. A., et al. 2005, PASP, 117, 721
\bibitem[\protect\citeauthoryear{Harris}{1996}]{1996AJ....112.1487H} Harris,W.~E., 1996, AJ, 112, 1487 
\bibitem[\protect\citeauthoryear{Impey, Wynn-Williams, \& Becklin}{1986}]{1986ApJ...309..572I} Impey C.~D., Wynn-Williams C.~G., Becklin E.~E., 1986, ApJ, 309, 572 
\bibitem[\protect\citeauthoryear{Irwin et al.}{2005}]{irwin2005} Irwin, M. J.,
Ferguson, A. M. N., Ibata, R. A., Lewis, G. F., \& Tanvir, N. R. 2005, ApJL, 
628, L105
\bibitem[\protect\citeauthoryear{Jeffery et al.}{2011}]{gulden2005}
  Jeffery, E. J., Smith, E., Brown, T. M., Sweigart, A. V., Kalirai,
  J., Ferguson, H. C., Guhathakurta, P., Renzini, A., \& Rich,
  R. M. 2011, AJ, 141, 171
\bibitem[\protect\citeauthoryear{Jensen et al.}{2003}]{jensen2003} Jensen, J. B., Tonry, J. L., 
Barris, B. J., Thompson, R. I., Liu, M. C., Rieke, M. J., Ajhar, E. A., \& Blakeslee, J. P. 2003, ApJ, 583, 712
\bibitem[\protect\citeauthoryear{Kalirai et al.}{2006}]{kalirai2006} Kalirai, J. S. et al. 2006, ApJ, 648, 389
\bibitem[\protect\citeauthoryear{Koch \& Grebel}{2006}]{kg2006} Koch, A., \& Grebel, E. K. 2006, 131, 1405
\bibitem[\protect\citeauthoryear{Kunder et al.}{2010}]{kund2010} Kunder, A., Chaboyer, B., \& Layden, 
A. C. 2010, AJ, 139, 415
\bibitem[\protect\citeauthoryear{Lauer et al.}{1998}]{lauer1998} Lauer, T. R., Faber, S. M.,
Ajhar, E. A., Grillmair, C. J., \& Scowen, P. A. 1998, AJ, 116, 2263
\bibitem[\protect\citeauthoryear{Martin et al.}{2006}]{martin2006} Martin, N. F. et al. 2006, MNRAS, 371, 1983
\bibitem[\protect\citeauthoryear{McConnachie et al.}{2005}]{mac2005} McConnachie, A. W., 
Irwin, M. J., Ferguson, A. M. N., Ibata, R. A., Lewis,  G. F., \& Tanvir, N. 2005, MNRAS, 356, 979
\bibitem[\protect\citeauthoryear{McConnachie et al.}{2009}]{mac2009} McConnachie, A. W. et al. 2009, 
Nature, 461, 66 
\bibitem[\protect\citeauthoryear{Monachesi et al.}{2011}]{Mona2011} Monachesi, A., Trager, S. C.,
Lauer, T. R., Freedman, W., Dressler, A., Grillmair, C., \& Mighell, K. J. 2011, ApJ, 727, 55
\bibitem[\protect\citeauthoryear{Monachesi et al.}{2012}]{Mona2012} Monachesi, A., Trager, S. C.,
Lauer, T. R., Hidalgo, S. L., Freedman, W., Dressler, A., Grillmair, C., \& Mighell, K. J. 2012, ApJ, 745, 97
\bibitem[\protect\citeauthoryear{Pritchet \& van den Bergh}{1987}]{PvdB1987} Pritchet, C. J. \& van den Bergh, S. 1987, ApJ, 316, 517
\bibitem[\protect\citeauthoryear{Pritchet \& van den Bergh}{1994}]{PvdB1994} Pritchet, C. J. \& van den Bergh, S. 1994, AJ, 107, 1730
\bibitem[\protect\citeauthoryear{Reiss \& Mack}{2004}]{reiss2004} Reiss, A., \& Mack, J. 2004, ISR-ACS 2004-06
\bibitem[\protect\citeauthoryear{Rose}{1985}]{rose1985}Rose, J. 1985, AJ, 90, 1927
\bibitem[\protect\citeauthoryear{Rose et al.}{2005}]{rose2005} Rose, J. A., Arimoto, N., 
Caldwell, N., Schiavon, R. P., Vazdekis, A., \& Yamada, Y. 2005, AJ, 129, 712
\bibitem[\protect\citeauthoryear{Rudenko et al.}{2009}]{rudenko2009} Rudenko, P., Worthey, G.,
Mateo, M. 2009, AJ, 138, 1985
\bibitem[\protect\citeauthoryear{Salaris \& Cassisi}{2005}]{sc2005}
  Salaris, M., Cassisi, S. 2005, Evolution of Stars and Stellar
  Populations (Chichester: Wiley)
\bibitem[\protect\citeauthoryear{Sarajedini \& Jablonka}{2005}]{sj2005} Sarajedini, A., \& Jablonka, P. 2005, AJ, 130, 1627
\bibitem[\protect\citeauthoryear{Sarajedini et al.}{2006}]{sara2006} Sarajedini, A., Barker, M., Geisler, 
D., Harding, P., Schommer, R. 2006, AJ, 132, 1361 
\bibitem[\protect\citeauthoryear{Sarajedini et al.}{2009}]{sara2009} Sarajedini, A., Mancone, C., Lauer, 
T. R., Dressler, A., Freedman, W., Trager, S. C. Grillmair, C., \& Mighell, K. J. 2009, AJ, 138, 184 (S09)
\bibitem[\protect\citeauthoryear{Schlegel, Finkbeiner, 
\& Davis}{1998}]{1998ApJ...500..525S} Schlegel D.~J., Finkbeiner D.~P., Davis M., 1998, ApJ, 500, 525 
\bibitem[\protect\citeauthoryear{Sirianni et al.}{2005}]{siri2005} Sirianni, M. et al. 2005, PASP, 117, 1049
\bibitem[\protect\citeauthoryear{Smith}{1995}]{sm1995} Smith, H. A. 1995, RR Lyraes, Cambridge 
Astrophysics Series, Vol. 27 (Cambridge, UK)
\bibitem[\protect\citeauthoryear{Stetson}{1987}]{stet1987} Stetson, P. B. 1987, PASP, 99, 191
\bibitem[\protect\citeauthoryear{Stetson}{1994}]{stet1994} Stetson,
  P. B. 1994, PASP, 106, 250
\bibitem[\protect\citeauthoryear{Suntzeff, Kinman, \&
    Kraft}{1991}]{1991ApJ...367..528S} Suntzeff, N.~B., Kinman, T.~D.,
  Kraft, R.~P., 1991, ApJ, 367, 528
\bibitem[\protect\citeauthoryear{Tammann, Sandage, 
\& Reindl}{2003}]{2003A&A...404..423T} Tammann, G.~A., Sandage, A., Reindl, B., 2003, A\&A, 404, 423 
\bibitem[\protect\citeauthoryear{Tonry et al.}{2001}]{tonry2001} Tonry, J. L., Dressler, A., 
Blakeslee, J. P., Ajhar, E. A., Fletcher, A. B., Luppino, G. A., Metzger, M. R., \& 
Moore, C. B. 2001, ApJ, 546, 681
\bibitem[\protect\citeauthoryear{Trager et al.}{2000}]{trager2000} Trager, S. C., Faber, S. M., 
Worthey, G., \& Gonzalez, J. J. 2000, AJ, 120, 165 
\bibitem[\protect\citeauthoryear{Worthey}{2004}]{worth2004} Worthey, G. 2004, AJ, 128, 2826 
\bibitem[\protect\citeauthoryear{Yang et al.}{2010}]{yang2010} Yang, S. -C., Sarajedini, A., 
Holtzman, J. A., \& Garnett, D. R. 2010, ApJ, 724, 799
\bibitem[\protect\citeauthoryear{Yang \& Sarajedini}{2012}]{yang2012} Yang, S. -C. \& Sarajedini, A. 2012, 
MNRAS, 419. 1362
\end{thebibliography}
\end{document}